\documentclass[12pt]{iopart}
\usepackage{xcolor}
\usepackage{caption}
\usepackage{subcaption}
\usepackage[english]{babel}
\usepackage[autostyle]{csquotes}
\usepackage{lipsum}
\usepackage{graphicx}
\usepackage{bm}
\usepackage{cite}
\usepackage[colorlinks, citecolor = blue, urlcolor = blue]{hyperref}
\def\be{\begin{equation}}
\def\ee{\end{equation}}

\begin{document}

\title{Memory of topologically constrained disorder in Shakti artificial spin ice}

\author{Priyanka Priyanka$^1$, Cristiano Nisoli$^2$ and Yair Shokef$^{1,3,4,5,6}$}

\address{$^1$ School of Mechanical Engineering, Tel Aviv University, Tel Aviv 69978, Israel}
\address{$^2$ Theoretical Division, Los Alamos National Laboratory, Los Alamos, New Mexico 87545, USA}
\address{$^3$ School of Physics and Astronomy, Tel Aviv University, Tel Aviv 69978, Israel}
\address{$^4$ Center for Computational Molecular and Materials Science, Tel Aviv University, Tel Aviv 69978, Israel}
\address{$^5$ Center for Physics and Chemistry of Living Systems, Tel Aviv University, Tel Aviv 69978, Israel}
\address{$^6$ International Institute for Sustainability with Knotted Chiral Meta Matter (WPI-SKCM$^2$), Higashi-Hiroshima, Hiroshima 739-8526, Japan}

\begin{abstract}
Complex behaviors often sit at a critical threshold between order and disorder. But not all disorder is created equal. Disorder can be trivial or constrained, and correlated disorder can even be topological. Crucially, constrained disorder can harbor memory, leading to non-trivial, sequence-dependent responses to external manipulations. And yet the fascinating subject of ``memory of disorder" remains poorly explored, as memory is often associated to the retention of metastable order. In recent years artificial frustrated materials---in particular arrays of frustrated nanomagnets known as artificial spin ices---have been employed to study complex disorders and its wealth of exotic behaviors, yet their memory properties have received much less attention. Here, we investigate both analytically and numerically the sequence-dependent responses of two somehow opposite yet related  artificial spin ices: the Landau-ordered square spin ice and the disordered but topologically-ordered Shakti spin ice. We find that Shakti exhibits a pronounced sequence-dependent response, whereas in the square lattice, such path dependence is absent. Within Shakti, even the minimal periodic supercell demonstrates both deterministic and stochastic forms of sequence memory, depending on the interaction strength. Extending our study to cyclic driving, we find that retracing the same input path leads to enhanced memory retention. These results open new perspectives on how topological constraints and correlated disorder generate robust memory effects in frustrated artificial materials, hitherto examined mainly in terms of their ground-state kinetics and thermodynamics.
\end{abstract}

\maketitle
\pagestyle{plain}
\tableofcontents
\clearpage
\pagestyle{plain}

\section{Introduction}
\label{sec:intro}

Complex systems such as granular matter~\cite{Lindeman2025}, foam~\cite{mukherji2019}, crumpled sheets~\cite{Shohat2022}, amorphous solids~\cite{fiocco2014, regev2015reversibility, mungan2019b, keim2020, keim2021multiperiodic, mungan2025}, origami~\cite{jules2022} and mechanical metamaterials~\cite{Ding2022, Sirote2024, Liu2024, Meulblok2025} typically accommodate multiple quasidegenerate or degenerate metastabilities often connected by constrained pathways~\cite{ritort2003glassy}, leading to memory in their response~\cite{keim2019}, which can explain biological function~\cite{mungan2025a} or could be exploited for technological applications~\cite{yasuda2021mechanical, paulsen2025a}. Perhaps the simplest and most notorious cases of memory-endowed response are provided by magnetism, where the simplest ferromagnet proves hysteretic. But more complex, disordered magnets, such as frustrated antiferromagnets, can reveal even more complex memory, including rate dependent memory~\cite{kudasov2006steplike}. Of particular interest are magnets that are topologically constrained~\cite{henley2010coulomb, henley2011classical}, whose phase space can be partitioned into sectors weakly connected by available pathways~\cite{Lieb1967, nisoli2020topological, zhang2023topological}. 

Recently, magnets with exotic behaviors have been designed with intention and realized at the nanoscale as arrays of ferromagnetic anisotropic nanoislands called artificial spin ices~\cite{Wang2006, nisoli2013}. The magnetic state of the nanoisland can be described as a binary, classical Ising spin which can be characterized individually through various techniques, sometimes even in real-time. Moreover, their flexible design allows for the deliberate generation of novel Hamiltonians to produce magnets of collective behaviors that are often absent in natural systems~\cite{nisoli2017deliberate, Sandra2020}. Among these intriguing properties, memory effects have received perhaps insufficient attention following some promising initial studies of direct visualization of avalanches of magnetic charges during moment inversion~\cite{Ladak2010, Mengotti2011, wysin2013dynamics}, and of return point memory~\cite{libal2012hysteresis, gilbert2015direct} related to the propagation of magnetic monopoles.

Here we investigate memory as sequence-dependent response in two widely studied artificial spin ices, which represent opposite extremes in their structural design and resulting low-energy manifolds: one is square ice, the first to be introduced nearly two decades ago~\cite{Wang2006, Nisoli2007}, and the other is Shakti spin ice~\cite{Chern2013, Morrison2013, gilbert2014emergent, lao2018classical, Stopfel2018, Carolina2023}, a more recent realization that belongs to a new generation of spin ices based on a distinct concept of vertex frustration~\cite{Morrison2013, stamps2014unhappy, gilbert2016emergent, Gilbert2016, nisoli2017deliberate, sultana2025}. Square spin ice~\cite{Wang2006, Nisoli2007, nisoli2010effective} obeys the ice rule, but its degeneracy is notoriously lifted, resulting in a conventional antiferromagnetic Landau order~\cite{morgan2011thermal, zhang2013crystallites, porro2013exploring} below a thermal phase transition of the Ising class~\cite{sendetskyi2019continuous}. In contrast, Shakti---derived by removing islands from square ice---exhibits a thermal crossover to a disordered phase characterized by algebraically correlated classical topological order~\cite{henley2011classical}. We intend that in the sense that its ground state can be mapped onto a dimer model~\cite{lao2018classical} or, equivalently, the F-model~\cite{Lieb1967} at the free fermion point~\cite{Chern2013}, both of which are inherently topological states. Moreover, in addition to topological order, for certain ratios among its couplings, Shakti also supports the trivial Landau order it inherits from square ice. Therefore, a comparative study of these two systems offers the potential to elucidate memory in ordered phases and in disordered, but topologically ordered, ones. Interestingly, we have previously demonstrated that both geometries can be adapted into mechanical metamaterials~\cite{coulais2016metacube, meeussen_NJP_2020, meeussen_NJP_2020, pisanty2021}, also resulting in intriguing forms of mechanical memory~\cite{merrigan2021topologically, Sirote2024}.

For both lattices, we consider athermal dynamics in which we quasistatically increase the in-plane magnetic field in different sequences. Initially, we focus on the ground-state configurations that exhibit the smallest repeating structures within the square and Shakti lattices, and we measure the magnetization at the conclusion of the process. In contrast with the simpler and predictable behavior that we find in square ice, for the more complex Shakti lattice, our study reveals a sequence-dependent response, which depends on the structural phase of the material. Moreover, we find that the response to applying an external field is not always deterministic; the dynamics can exhibit bifurcations, leading to multiple possible final states.

Whether the system can return to the ground state configuration when the magnetic field is switched off depends on factors such as spin orientation, the locations of defects (or excitations) where the vertices are not in their minimum energy configurations, and the selected protocols for the evolution of the external magnetic field. However, in large systems, regardless of the protocols employed, the system does not revert to its initial state once it is driven beyond a certain level. This phenomenon occurs because large systems are composed of overlapping supercells, which results in a reduction of the external field threshold to the minimum threshold observed in periodic arrangements of the individual supercells.

We chose and followed protocols that can be replicated in an experimental setting with magnetic force microscopy~(MFM) under field. Our results might prove relevant in the context of magneto-transport in these systems~\cite{branford2012emerging, chern2017magnetotransport, le2017understanding, caravelli2022artificial}. More broadly, we envision a series of experiments in which a small sized, static Shakti (that is an artificial spin ice with nanoislands of circa 20~nm of thickness),  can be repetitively subjected to applied field and visualized via e.g. MFM after each application.

In Sec.~\ref{sec:model} we present the two lattices and the various protocols that we employed for studying their responses to changing the external magnetic field. In Sec.~\ref{sec:square} we analyze the response of the square lattice to these protocols. In Sec.~\ref{sec:Shakti1D} we study the response of the Shakti lattice to an external field in a single direction, and in Sec.~\ref{sec:Shakti2D} we study its response to protocols in which both components of the in-plane external field are varied. Section~\ref{sec:conclusions} provides conclusions and discussion of the results and their implications for further research.

\begin{figure*}[t]
\centering
\includegraphics[width=\columnwidth]{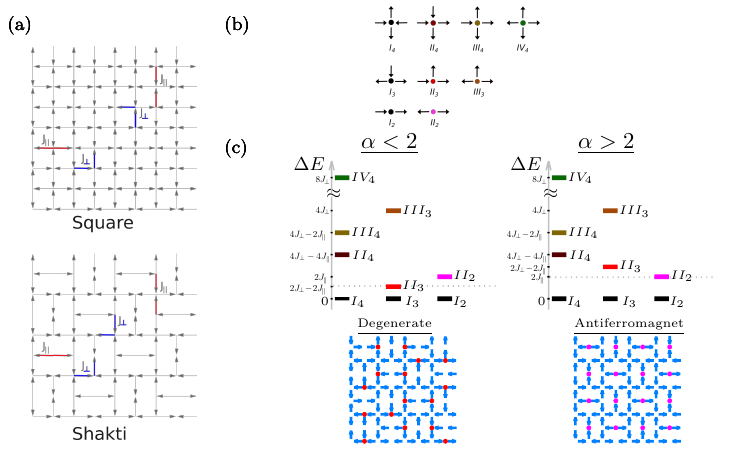}
\caption{a) Dilution of the square lattice (top) yields the Shakti lattice (bottom). Representative perpendicular (blue) and parallel (red) interacting spin pairs are marked. b)~Spin configurations of ascending energy, in four-, three- and two-coordinated vertices. c)~Energy spectra of vertex types (top) and ground state configurations (bottom) for the Shakti lattice. Energy is plotted with respect to the minimal energy level for each coordination number. Colors correspond to vertex colors in b. The lattice does not permit all vertices to be in their lowest energy state, and the first excitation switches from $II_3$ to $II_2$ as the ratio of perpendicular to parallel interaction $\alpha = J_{\perp} / J_{\parallel}$ changes, leading to a degenerate ground state for $\alpha<2$ (left) and to an antiferromagnetic ground state for $\alpha>2$ (right).}
\label{lattice}
\end{figure*}

\section{Model}
\label{sec:model}

\subsection{Geometries}

\paragraph{Square Lattice.} It is convenient to start with square ice with periodic boundary conditions since the Shakti lattice can be obtained from it by ordered decimation. Square ice, (Fig.~\ref{lattice}a, top) is composed of binary magnetic moments (describing magnetic nanoislands) pointing along the edges of a square lattice and impinging in the vertices. The moments interact as dipoles, but we will employ a ``tried and true"~\cite{Wang2006, Nisoli2007, nisoli2010effective} nearest neighbor approximation where $J_{\parallel}$ and $J_{\perp}$ are the couplings between parallel and perpendicular islands impinging in the same vertex. Thus the system is governed by the Hamiltonian,
\be
\mathcal{H} = -\displaystyle{J_{\parallel}} \sum_{{\langle ij \rangle}_{\parallel}} \vec{S}_i\cdot\vec{S}_j-{J_{\perp}}\sum_{{\langle ij \rangle}_{\perp}}(\vec{S_i}\times\vec{S_j})\cdot \hat{z}+\vec{h}\cdot \sum_i \vec{S}_i~,
\label{Hami}
\ee
where $\vec{h} = (h_x, h_y)$ represents the uniform external field applied to all the spins in the lattice. The angular brackets $\langle i j \rangle$ denote nearest-neighbor spins that share a vertex, and the summation is taken over all nearest-neighbor pairs. In what follows we measure interaction strengths and external fields in units in which $J_{\parallel}=1$, and the system's response will depend on the dimensionless ratio $\alpha=J_{\perp}/J_{\parallel}$ between the two interaction strengths.

For the square lattice, if $\alpha=1$ the ground state becomes the celebrated six-vertex model~\cite{lieb1967residual}, a disordered tesselation of ice-rule-obeying vertices (two spins pointing in, two out) which realizes a classical topological phase~\cite{henley2010coulomb, nisoli2020topological}. In general, banning specific tricks in fabrication~\cite{moller2009magnetic, perrin2016extensive, schanilec2022approaching}, the couplings inherited from the dipolar interaction correspond to $\alpha>1$ leading to a trivial antiferromagnetic ground state~\cite{morgan2011thermal, zhang2013crystallites, porro2013exploring} shown in Fig.~\ref{lattice}a, top.

To connect the square ice to celebrated vertex models, it is customary~\cite{Wang2006, Nisoli2007, nisoli2010effective} to describe it in terms of the energies of its vertex topologies, shown in Fig.~\ref{lattice}b in order of increasing energy (for the case $\alpha>1$), and denoted as Type $I_4$ to $IV_4$. A moment's thought shows that vertices are frustrated: no configurations can minimize all spin pairs. Yet, for $\alpha>1$ the system is ordered, as shown in Fig.~\ref{lattice}a, top: Type $I_4$ and Type $II_4$ vertices obey the ice rule (two moments in, two out) but Type $I_4$ has lower energy than Type $II_4$. The energy difference between vertices is shown in the four-coordinated vertex column of Table~\ref{lattice}. We stress that these considerations are valid for the case $\alpha>1$, the only case considered here. 

\begin{table}[h]
\centering
\begin{tabular}{|l|l|l|}
\hline 
$E(IV_4) = 2 J_{\parallel} + 4 J_{\perp} = 2 + 4 \alpha$ & $E(III_3) = J_{\parallel} + 2 J_{\perp} = 1 + 2 \alpha$ & $E(II_2) = J_{\parallel} = 1$ \\
$E(III_4) = 0$ & $E(II_3) = -J_{\parallel} = -1$ &  $E(I_2) = -J_{\parallel} = -1$ \\
$E(II_4) = -2 J_{\parallel} = -2$ & $E(I_3) = J_{\parallel} -2 J_{\perp} = 1 - 2 \alpha$ & \\
$E(I_4) = 2J_{\parallel} -4J_{\perp} = 2 - 4 \alpha$ & &  \\
\hline
\end{tabular}
\caption{Energies of all vertex types. Working in units in which $J_\parallel=1$ allows to write all vertex energies in terms of the dimensionless ratio $\alpha = J_\perp/J_\parallel$.}
\label{energy}
\end{table}

\paragraph{Shakti Lattice.} The Shakti artificial spin ice, as illustrated in Fig.~\ref{lattice}a, bottom, is obtained by the alternate removal of two vertically or horizontally aligned spins at every interval of four spins, leading to vertices of coordination 4, 3, and 2. Here too one can proceed by a vertex description, and Fig.~\ref{lattice}b shows the vertex topologies for different coordination, and Table~\ref{lattice} shows their energies expressed in terms of $J_{\parallel}$ and $J_{\perp}$, and then simplified in the dimensionless units that we use in this paper.

For $\alpha<2$, not all vertices of Shakti can be placed in their lowest energy configuration. This is a case of so-called ``vertex-frustration'' where disorder and degeneracy come not from frustration among the spins in a single vertex but rather the allocation of vertex configurations~\cite{Morrison2013, Chern2013}. This new notion of frustration has led to a variety of new nano-fabrications of distinct exotic behavior~\cite{nisoli2017deliberate} and it has been shown to map into the concept of incompatibility in metamechanics~\cite{Sirote2024}. Then, the lowest energy configuration is shown in Fig.~\ref{lattice}c, left: where some of the vertices of coordination 3 are forced by the geometry to be in the excited state $II_3$. We have shown that the allocation of these ``unhappy vertices'' maps into a particular six-vertex model, the F-model~\cite{Chern2013} but also into the dimer cover model on a square lattice~\cite{lao2018classical}, both of which realize topological phases. 

However, when $\alpha>2$ the parallel coupling is weak enough that it is energetically convenient to put all the unhappy vertices on vertices of coordination 2, leading to an ordered antiferromagnetic configuration {\it which is simply the decimated ground state of square ice, from which Shakti has been obtained}. This can be understood by looking at the energy spectra in Fig.~\ref{lattice}c, showing that the first excitation is on vertices of coordination 3 when $\alpha<2$, while when $\alpha>2$ it is on vertices of coordination 2.

\subsection{Probe-Response Protocols}

We investigate protocol dependent response when an external magnetic field is applied either to square ice starting in its antiferromagnetic ordered ground state, or to Shakti ice for the interaction ratio, $\alpha$ larger and smaller than 2, and starting in both states, the disordered, topological one and the antiferromagnetic one. We change the external field quasistatically, namely very slowly over a long enough time interval~$\tau$, so that the system constantly remains in a local minimum energy state. To characterize the system's state, we extract its average in-plane magnetization, 
\be
\vec{M} = \frac{1}{L^2} \sum_i \vec{S_i}.
\ee

\begin{figure}[t]
\centering
\includegraphics[width=\columnwidth]{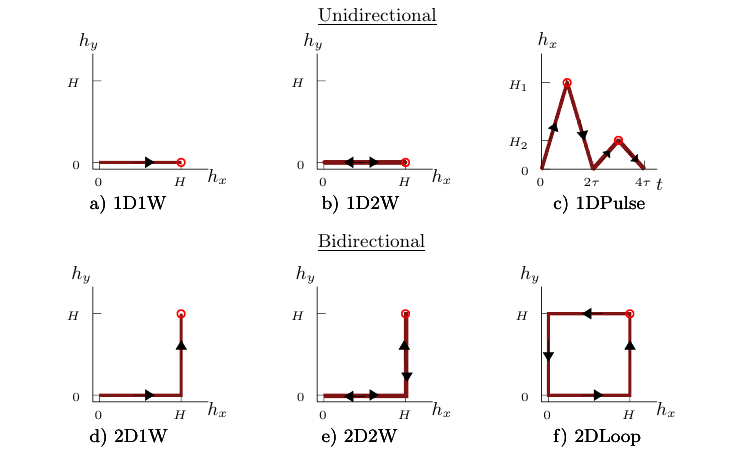}
\caption{The protocols used to drive the artificial spin ice by an in-plane external magnetic field $\vec{h}=(h_x,h_y)$. Unidirectional protocols: a) One way (1D1W) X path; b) Two way (1D2W) XX path; c) Pulse (1DPulse) XX path. Bidirectional protocols: d) One way (2D1W) XY path; e) Two way (2D2W) XYYX path; f) Loop (2DLoop) XYXY path. All six protocols have a complementary variant in which X and Y are exchanged. Red dots indicate the maximal external fields applied during the process.}
\label{Prtime}
\end{figure}

We consider different protocols for applying an in-plane external magnetic field. These are illustrated in the \((h_x, h_y)\)-plane in Fig.~\ref{Prtime}:

\begin{enumerate}
\item \textbf{Unidirectional (1D) protocols} -- The magnetic field is applied along a single direction:
\begin{itemize}
\item \emph{One-way} (1D1W): The field is increased monotonically to a target value \( H \) (Fig.~\ref{Prtime}a).
\item \emph{Two-way} (1D2W): The field is ramped up to \( H \), then reduced back to zero (Fig.~\ref{Prtime}b).
\item \emph{Pulse} (1DPulse): Two successive pulses of different strengths \( H_1 \) and \( H_2 \) are applied, each followed by a return to zero field (Fig.~\ref{Prtime}c).
\end{itemize}

\item \textbf{Bidirectional (2D) protocols} -- The magnetic field is applied in both in-plane directions:
\begin{itemize}
\item \emph{One-way} (2D1W): The field is increased first along one axis, then along the other, to finally reach \( \vec{h} = (H, H) \). First increasing $h_x$ and then increasing $h_y$ is denoted XY (Fig.~\ref{Prtime}d), while increasing the two components of $\vec{h}$ in the oppositve order is denoted YX.
\item \emph{Two-way} (2D2W): After increasing both components of the field, the field returns to zero along the same path used to reach \( \vec{h} = (H , H) \), namely XYYX (Fig.~\ref{Prtime}e) or alternatively YXXY.
\item \emph{Loop} (2DLoop): After following one path to \( \vec{h} = (H, H) \), the field is removed along the reverse path, namely XYXY (Fig.~\ref{Prtime}f) or YXYX.
\end{itemize}
\end{enumerate}

For the various protocols, we analyze how the final state depends on the sequence of modifying the external field and on the maximum field strength $H$. Specifically, protocol 1D2W is the simplest situation in which we can ask whether the system's state depends on its history. For protocol 1DPulse we explore how the final state depends on the order of the two pulses. For protocol 2D1W, we examine sequence dependence by comparing XY and YX sequences that end at the same external field. For the 1D2W, 2D2W and 2DLoop protocols that end at zero field, we ask whether the system ends in its initial configuration, or in some other ground state configuration, or possibly in an excited state, which is thus metastable.

For small systems, we analytically track the dynamics to flip the spins, whereas for large systems, we perform zero-temperature single-flip Monte Carlo simulations using the Metropolis algorithm. For the Shakti lattice, Instead of using the thermalization method to generate ground state configurations~\cite{morgan2011thermal, Kapaklis2012}, we numerically generate the ground state following the order of defects such that the spins also follow the ice rule constraint. In this method, we randomly assign the positions of the three-coordinated vertices, ensuring that each cell has two defect states ($II_3$) and two non-defect states ($I_3$). Once we have the base configurations with three-coordinated vertex defect states ready, we assign the spins in a way that all four-, two-coordinated, and the remaining three-coordinated vertices are in their minimum energy (non-defect) states, $I_4$, $I_2$, and $I_3$, respectively. 

\section{Response of Square Artificial Spin Ice}
\label{sec:square}

\begin{figure}[h!]
\centering
\includegraphics[width=\columnwidth]{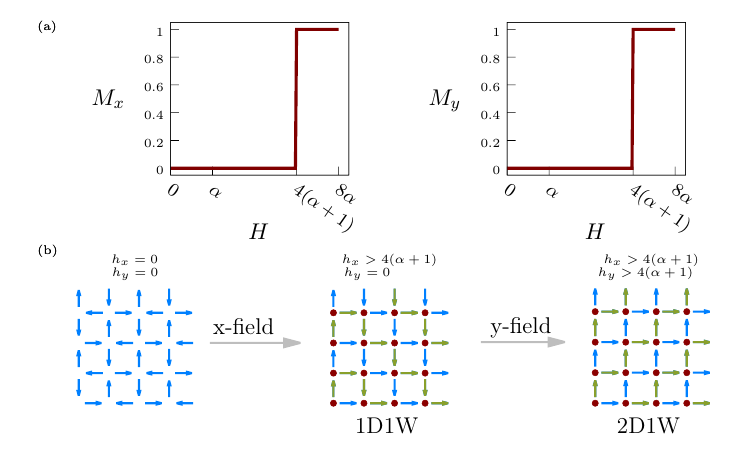}
\caption{Response of square artificial spin ice to one-way protocols. a) Magnetization at the end of the 1D1W or 2D1W protocols as a function of the final field. b) Left: Ground state configuration at zero field. Center: Representative configuration after 1D1W, namely at field $\vec{h}=(H,0)$, where $H$ is greater than the critical field $H_c=4(\alpha+1)$; all $x$-spins point along the field, while $y$-spins are arranged such that each column points in an arbitrary direction. Right: Configuration at the end of the 2D1W protocol, namely at the final field $\vec{h}=(H,H)$. The flipped spins with respect to the initial configuration are colored in green, and excited vertices are colored according to the scheme introduced in Fig \ref{lattice}b.}
\label{Squaremag}
\end{figure}

We begin by examining the response to an external magnetic field of the square lattice in its antiferromagnetic ground state (Fig.~\ref{lattice}a, top). At zero temperature, spins can flip when the field exceeds $H_c = 4J_\perp +4J_\parallel = 4(\alpha+1)$, using our convention that $J_\parallel=1$. This value of~$H_c$ corresponds to the Zeeman energy required to excite the lowest-energy four-coordinated spin configuration Type~$I_4$ to the transient Type~$III_4$ by a single spin flip. Then, vertices fall back into Type~$II_4$. Thus, the magnetization as function of the external field sharply increases at~$H_c$ from~0 to~1, as shown in Fig.~\ref{Squaremag}a.

We first consider the system's response to a quasistatic increase of the magnetic field in the $x$-direction, namely following protocol~1D1W. As the field reaches the threshold~$H_c$, vertices transition to the excited~$II_4$ state. A single spin flip triggers a cascade, aligning all spins with the field and minimizing the energy. The system exhibits~$2^L$ degenerate configurations, where all~$x$-spins align with the field, and~$y$-spins arrange such that all vertices are of Type~$II_2$, but each column can independently orient along the positive or negative~$y$-direction. Thus generating a partially disordered state with sub-extensive degeneracy, see Fig.~\ref{Squaremag}b, center. The bidirectional one-way protocol, 2D1W continues from this state, and as the field in the $y$-direction exceeds the same threshold $H_c$, all $y$-spins align with the field, as shown in Fig.~\ref{Squaremag}b, right. 

The YX bidirectional one-way protocol 2D1W ends in the same final state as protocol XY, discussed so far, with an analogous intermediate state, with $x$ and $y$ exchanged. Due to the ordered and isotropic nature of the square lattice, the partially-disordered intermediate state leaves no memory imprint: as the field increases beyond $H_c$, all spins align with the field, and no sequence dependence is observed. 

For the two-way and loop protocols, the system stays in the ground state if the applied field remains below the critical threshold~$H_c$. When the field exceeds~$H_c$, the final state consists of all spins aligned with the field direction, forming a uniform Type~$II_4$ configuration. This behavior is independent of the interaction strength ratio~$\alpha$.

\section{Response of Shakti Artificial Spin Ice to Unidirectional Protocols}
\label{sec:Shakti1D}

\subsection{Degenerate Ground State of Shakti Artificial Spin Ice}
\label{sec:degenerate-states}

The behavior of the Shakti lattice is considerably more complex than that of the square lattice. We start by analyzing the ground state configurations on it, classifying them by their symmetry and magnetization. The $4 \times 4$ Shakti \emph{supercell} is the smallest block that may be repeated to form this lattice. We begin by considering spin configurations on the supercell, which may in turn be repeated to form periodic spin states for an extended lattice. Subsequently, we will also consider lattices of arbitrary size $L \times L$, with a spin state which does not necessarily have this periodicity, namely that different supercells within it may have different spin configurations.

For $\alpha > 2$, the lowest excited vertex is Type~$II_2$. Therefore, the ground state of Shakti is the ordered antiferromagnetic state (Fig.~\ref{lattice}c), with all 4-coordinated vertices of Type~$I_4$, all 3-coordinated vertices of Type~$I_3$, and the frustration is resolved by the 2-coordinated vertices, which are all excited to Type~$II_2$, and may be thought of as the defects in the spin configuration.

For $\alpha < 2$, the lowest excited vertex is Type~$II_3$. Therefore, the defects concentrate on the 3-coordinated vertices; the ground state supercell has all 4-coordinated vertices of Type~$I_4$, all 2-coordinated vertices of Type~$I_2$, while four of the 3-coordinated vertices are of Type~$I_3$, and the remaining four are of Type~$II_3$. By parity of spin flips, and due to the periodic boundary conditions, each row or column in the supercell can have either zero or two such Type~$II_3$ defects. Nonetheless, this minimal supercell has twenty-four different spin configurations that minimize the energy, and these constitute what we will refer to as the \emph{degenerate ground state}.

\begin{figure}[t]
\centering
\includegraphics[width=0.8\columnwidth]{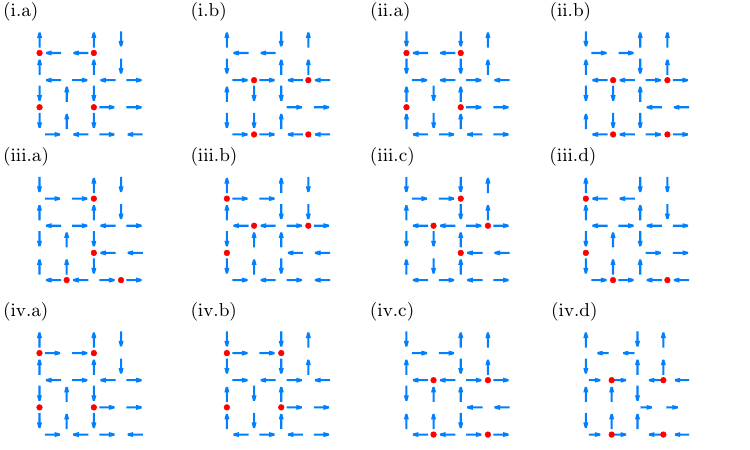}
\caption{Degenerate ground state configurations of the Shakti supercell. Excited 3-coordinated vertices are colored according to the scheme introduced in Fig.~\ref{lattice}b. Configurations (i)-(iii) all have zero magnetization, while configurations (iv.a) and (iv.b) have positive $x$-magnetization and configurations (iv.c) and (iv.d) have positive $y$-magnetization.}
\label{Configl4}
\end{figure}

These may be reduced to twelve, modulo the $Z_2$ symmetry which flips all the spins, but does not change the allocation of the unhappy vertices. The reduced set of twelve configurations is further partitioned into four types based on magnetization; As illustrated in Fig.~\ref{Configl4}, all the configurations in groups (i) to (iii) have zero magnetization in both the $x$ and $y$ directions, whereas the configurations in group (iv) are magnetized. 

Considering an extended lattice made of these supercells, or equivalently these supercells with periodic boundary conditions, the configurations within the group are classified based on their rotational and translational symmetry, with identical response to the application of an external field. Notably, group (iii) contains four configurations that in an extended lattice are equivalent by translational symmetry. 

Finally, we denote by \(\ ( \overline{\textrm{i}}) \) to \( (\overline{\textrm{iv}}) \) the complementary configurations that are obtained from configurations (i) to (iv) by flipping all spins in the supercell.

Despite the distinction in the ground state as $\alpha$ is varied -- namely degenerate for $\alpha < 2$ and antiferromagnetic for $\alpha > 2$ -- in the absence of an external field, both states are locally stable. Namely, any single spin flipped will increase the energy. Therefore, when analyzing the response of Shakti artificial spin ice to different protocols of modifying the external field, we will consider initial states that are either the antiferromagnetic or the degenerate ones, irrespective of the value of $\alpha$, namely both for starting in the ground state and for starting in a metastable state.

\begin{figure}[t]
\centering
\includegraphics[width=\columnwidth]{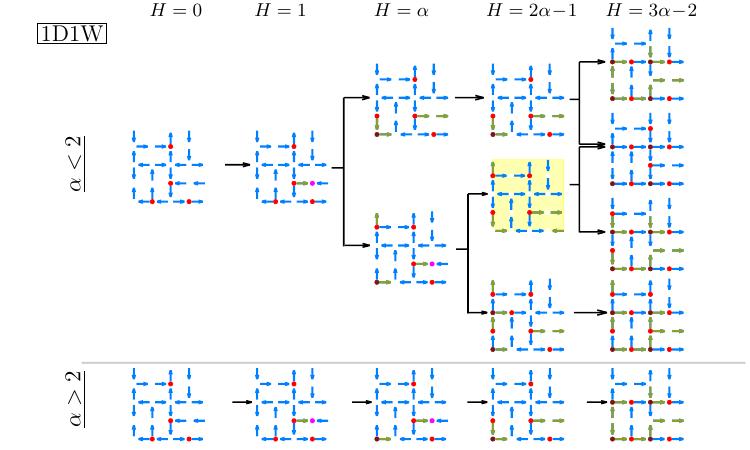}
\caption{Response starting from the degenerate state. Minimum energy states obtained at the end of the one way unidirectional protocol 1D1W, starting at configuration (iii.a). The $\alpha<2$ case (top) shows bifurcations in the outcomes for $H>\alpha$ whereas for $\alpha>2$ (bottom), the response is deterministic. The degenerate ground state configuration (iv.a) is highlighted in yellow.}
\label{Uni_DGS12}
\end{figure}

Unlike square ice, in Shakti we see non-deterministic behavior where we observe non-unique avalanches of spin flips that lead to multiple possible final configurations. As detailed below, this can happen for any value of $\alpha$, and starting both from the structurally degenerate state or from the antiferromagnet initial state. In this section we study the response of the Shakti lattice to the different unidirectional protocols defined above, and starting either at the degenerate state or at the antiferromagnetic state. We will see that the responses can differ depending on the value of $\alpha$, namely when $\alpha<2$, where the ground state is the degenerate one vs. when $\alpha>2$ and the ground state is antiferromagnetic, and with possibly singular behavior at $\alpha=2$. Section~\ref{sec:Shakti2D} presents similar analysis for the different bidirectional protocols with the Shakti lattice.

\subsection{One Way Protocol, 1D1W}

\paragraph{Degenerate state:} In unidirectional protocols, there is no sequence dependence since the external field is applied in only one direction, either along the \(x\) or \(y\) axis. Starting from the degenerate state, for \(\alpha \ge 2\), we observe deterministic dynamics where the magnetization aligns with the applied external field, and the magnetization in the complementary direction remains unchanged from its initial value. However, for \(\alpha < 2\) (see Fig.~\ref{Uni_DGS12}, top), the magnetization in the direction complementary to the applied field does change. This flipping of multiple spins leads to bifurcations and multiple stable states as outcomes. The critical value of the external field depends on the specific initial configuration and is in the range $1 < H_c \le 3\alpha-2$.

\begin{figure}[t]
\centering
\begin{subfigure}{0.9\columnwidth}
\includegraphics[width=\columnwidth]{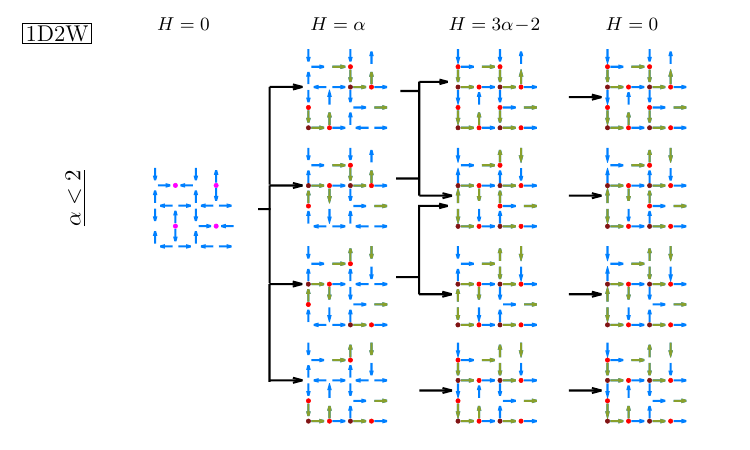}
\end{subfigure}
\begin{subfigure}{0.9\columnwidth}
\includegraphics[width=\columnwidth]{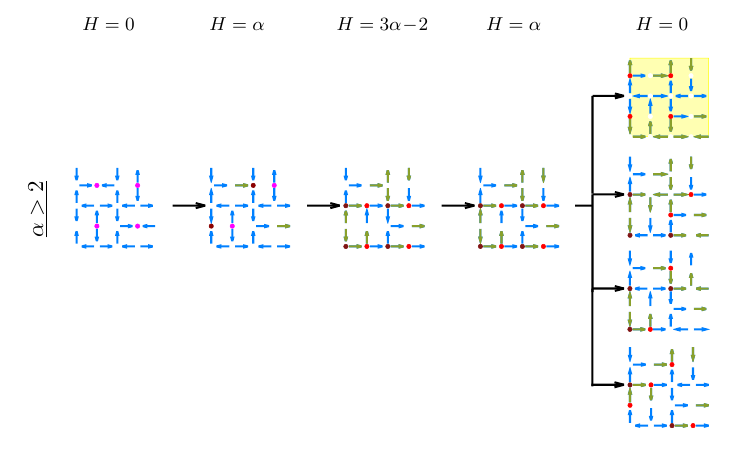}
\end{subfigure}
\caption{Response starting from the antiferromagnetic state. Minimum energy states obtained in the unidirectional two-way protocol 1D2W starting in the antiferromagnet configuration. For $\alpha\le2$ (top), There are multiple possible outcomes obtained for $H>\alpha$. Interestingly, turning off the external field to $H=0$ from the maximal external field $H=3\alpha-2$, the system remains in the same excited state. For $\alpha>2$ (bottom), the dynamics are deterministic for increasing the magnetic field, but while decreasing the field, there are multiple possible outcomes, and one of the configuration obtained is the degenerate state (iv.a) highlighted in yellow.}
\label{AF12wayConfig}
\end{figure}

\paragraph{Antiferromagnet state:} When a unidirectional external field is applied to the antiferromagnetic configuration, the outcomes depend on the parameter \(\alpha\), as demonstrated in Fig.~\ref{AF12wayConfig}. For \(\alpha > 2\), the spins in the direction of the external field align such that four- and three-coordinated defects are positioned in rows after reaching the threshold field $H_c = 3\alpha - 2$. Conversely, for $\alpha \le 2$, all vertices become excited at this same threshold~$H_c$ of the external field. Furthermore, we observe bifurcations at a higher value of the external field for all values of $\alpha$; the external field threshold for $\alpha>2$ is much larger than for $\alpha\le 2$. 

\subsection{Two Way Protocol, 1D2W}

\paragraph{Degenerate state:} For \(\alpha > 2\), when returning to zero magnetic field using the two-way protocol, we observe that in the presence of a small field, the system reverts to its initial configuration. Furthermore, even for very large fields, the system typically returns to one of the degenerate configurations, but to one with larger magnetization than the initial state. In other words, the field takes the system to an excited magnetized configuration, and when it is removed the system falls back into the magnetized part of the degenerate state.  

\begin{figure}[t]
\centering
\includegraphics[width=\columnwidth]{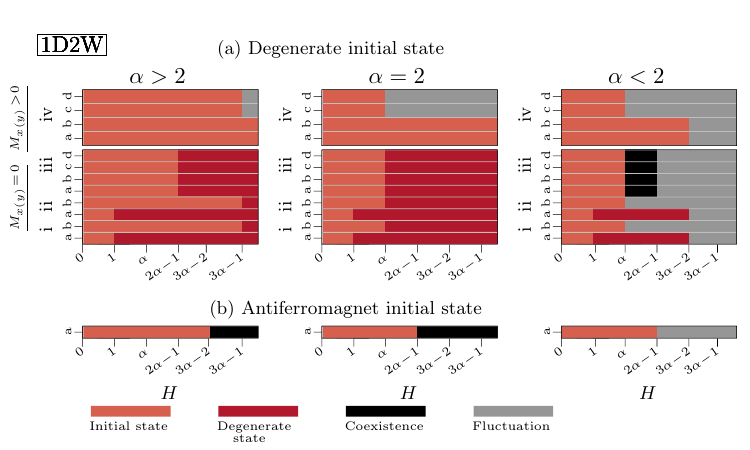}
\caption{Memory retention at the end of the unidirectional two-way protocol 1D2W as function of the maximal external field $H$ for different ranges of $\alpha$, starting at all degenerate states (a) or at the antiferromagnetic state (b). Colors indicate whether the system ends in its initial state (orange) or in some other degenerate state (red), may coexist between various possible excited or degenerate states (black), or fluctuates between various excited states (dark gray).}
\label{UniD}
\end{figure}

Notably, for the XX path, except for configuration (iv.c) and (iv.d) that have finite \(y\)-magnetization, all other configurations among the degenerate state revert to another degenerate state with positive \(x\)-magnetization and zero \(y\)-magnetization, as illustrated in the first column of Fig.~\ref{UniD}a. This suggests that the system does not retain any memory of its initial state and prefers to switch to a state with higher magnetization. We observe the same qualitative behavior for $\alpha=2$. For configuration (iv.c) and (iv.d), after applying a very high field the system outcome fluctuates between different excited states (shown by the dark gray bar in Fig. \ref{UniD}).

Instead, for \(\alpha < 2\), once the system becomes excited, it loses the ability to return to any degenerate state, as seen in Fig.~\ref{UniD}. Configuration~(iii) which possesses symmetry in the extended lattice exhibits coexistence (shown by the black bar in Fig.~\ref{UniD}) due to bifurcations during the process of reaching the maximal external field~(see Fig.~\ref{Uni_DGS12}). Here, when returning to zero field, some configurations get stuck in a local energy minimum and remain in an excited state, while other configurations return to degenerate Shakti configurations that belong to group~(iii).

\begin{figure}
\centering
\includegraphics[width=\columnwidth]{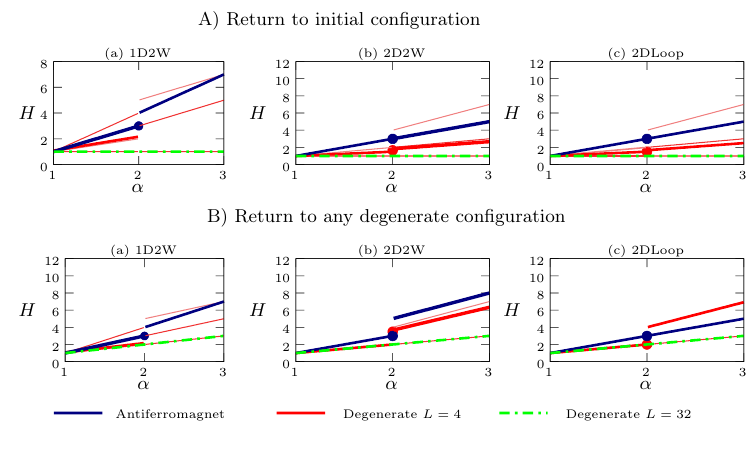}
\caption{Memory retention threshold field as a function of the interaction ratio $\alpha$ for different protocols: a) Unidirectional two way, b) Bidirectional two way, and c) Bidirectional Loop. Red and blue lines represent the cutoff field for the $4 \times 4$ supercell, whereas the green line represents the average cutoff for a lattice of size $L=32$. Solid thick lines represent the average cutoff magnetic field after which the $L=4$ supercell does not return to any of the degenerate states, whereas thin lines are the cutoffs for the individual configurations for $L=4$. The broken line is the average cutoff for $L=32$. Solid dots are the average cutoff values of the field at $\alpha=2$.}
\label{cutoff}
\end{figure}

Additionally, we analyze the average threshold field required to return to the same configuration as a function of the parameter \(\alpha\). We find a discontinuity in $H_c$ vs $\alpha$ at \(\alpha = 2\). For \(\alpha > 2\), the average threshold becomes infinite due to configuration (iv), which consistently reverts to its same configuration even at very high fields, as shown in Fig.~\ref{cutoff}a.

\paragraph{Antiferromagnet state:} Antiferromagnetic configurations exhibit stochastic behavior when returning to zero magnetic field using a two-way protocol. For \(\alpha \ge 2\), the final configuration is either one of the degenerate states with higher magnetization (group iv) or failure to return to the degenerate state (see Fig.~\ref{AF12wayConfig}). Conversely, for \(\alpha < 2\), the outcomes are stochastic, and the system ends in one of the excited configurations, and never returns to the degenerate state. This result is also illustrated in Fig.~\ref{UniD}b. The threshold field required to return to the same ground state configuration is represented by a straight line, with different slopes for \(\alpha < 2\) and \(\alpha > 2\), exhibiting a discontinuity at \(\alpha = 2\). The slope for \(\alpha > 2\) is greater than that for \(\alpha < 2\), see Fig~\ref{cutoff}a. This larger field threshold at \(\alpha > 2\) may be a characteristic that arises from the stability of the antiferromagnetic ground state.

\subsection{Pulse Protocol, 1DPulse}

\begin{figure}[t]
\centering
\includegraphics[width=\columnwidth]{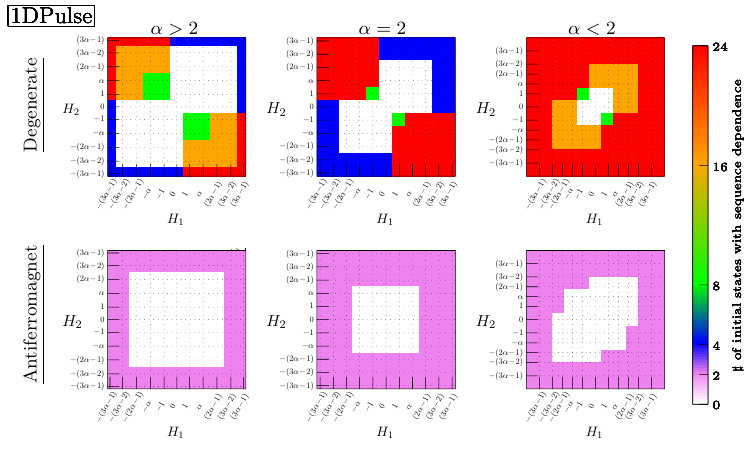}
\caption{Number of initial configurations with sequence dependence after the unidirectional pulse protocol 1DPulse, presented as function of the magnitudes of the two pulses. Results are shown starting in the degenerate state (top) or in the antiferromagnetic state (bottom), and for different ranges of the interaction ratio $\alpha$.}
\label{HisPulse}
\end{figure}

Sequence dependent response in unidirectional protocols may be considered using the pulse protocol. We need a two-dimensional system, either in time or space, to break symmetry and allow for the sequence response. When the amplitudes of the two pulses are equal and in the same direction, we do not observe any sequence-dependence response, as there is no asymmetry present. However, when the amplitudes differ, the sequence dependence response is influenced by the parameter \(\alpha\) for both the degenerate and the antiferromagnetic states, as described below.

\paragraph{Degenerate State:} For \(\alpha > 2\), sequence dependence becomes noticeable when the amplitude of one of the pulses is sufficiently large. Specifically, if one pulse is strong enough to effectively align nearly all spins in the direction of the applied field, we can observe this sequence-dependent response. In Fig.~\ref{HisPulse}, we show the total number of initial configurations displaying sequence-dependent response as function of the amplitudes of the two pulses. The empty region indicates the absence of any sequence dependent response, which is particularly prominent for \(\alpha > 2\). In this case, we observe a maximum of eight configurations out of twenty-four showing this response. In contrast, for \(\alpha < 2\), once a certain threshold in the field is reached, all configurations exhibit sequence-dependent response. Furthermore, the final configuration at the end of the protocol is not unique. This implies a stochastic response due to bifurcations, which becomes more pronounced at higher external fields.

\paragraph{Antiferromagnet ground state:} Similarly to the degenerate state, sequence dependence for \(\alpha > 2\) requires a significantly larger difference between the two pulses compared to the case where \(\alpha \leq 2\). The response includes stochasticity for all values of $\alpha$.

\section{Response of Shakti Artificial Spin Ice to Bidirectional Protocols}
\label{sec:Shakti2D}

We now consider the response of the Shakti lattice to the various bidirectional protocols, in which the external field includes components in both in-plane directions. As with the unidirectional cases studied above, also here we consider initial states that are either from the degenerate state or from the antiferromagnetic state, and we observe that the results differ depending on whether $\alpha < 2$ or $\alpha > 2$.

\subsection{One Way Protocol, 2D1W}

\begin{figure}[t]
\centering   \includegraphics[width=0.9\columnwidth]{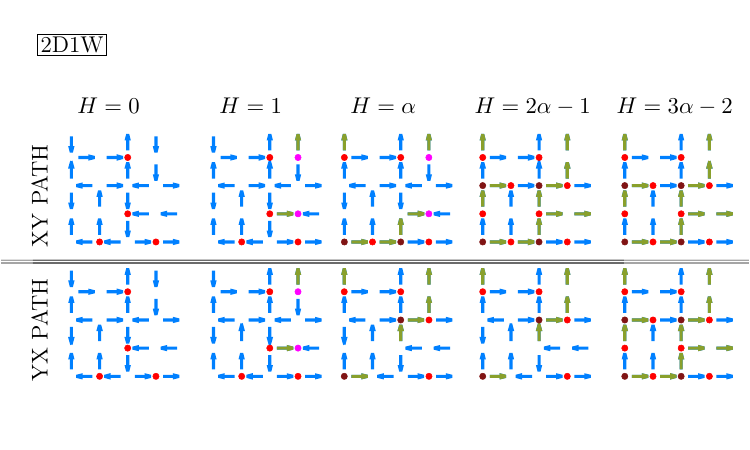}
\caption{Response of the degenerate state to the bidirectional one way protocol. Outcomes of the XY and YX paths at the end of the bidirectional one way protocol starting from configuration (iii.a) for $\alpha>2$ and with varying values $H$ of the final field. At intermediate values $\alpha<H<3\alpha-2$, the outcome is path dependent whereas for small field the system remains in its initial state. For extremely large field $H>3\alpha-2$, all spins flip in the direction of the field in both cases. The flipped spins with respect to the initial, zero-field configuration are colored in green, and excited vertices are colored according to scheme introduced in Fig.~\ref{lattice}b.}
\label{DFGS02}
\end{figure}

\paragraph{Degenerate state:} Unlike the unidirectional protocols, the bidirectional one-way protocol affects the flipping of spins in both $x$ and $y$-directions and causes asymmetry with the difference in the direction of the applied field. Thus, we expect sequence-dependence response in this case. We begin by analyzing the effect of an external magnetic field on the super-cell of Shakti for \(\alpha > 2\) using the one-way protocol. The sequence-dependent response occurs when either \(x\)- or \(y\)-spins are flipped at a specific external magnetic field. When the magnetic field is very weak (\(H \leq 1\)), the initial alteration in the energy landscape involves flipping a spin that connects the two-coordinated and three-coordinated vertices. This means either the \(x\) or \(y\) spin can flip without impacting the orthogonal spins shared by the same vertex. Consequently, the outcome is independent of the sequence or path, as shown in Fig.~\ref{DFGS02}. It is also noteworthy that a spin can flip at a field \(H \leq 1\) only when it connects three-coordinated vertices of Types \(II_3\) and \(I_3\). 

\begin{figure}[t!]
\centering\includegraphics[width=\columnwidth]{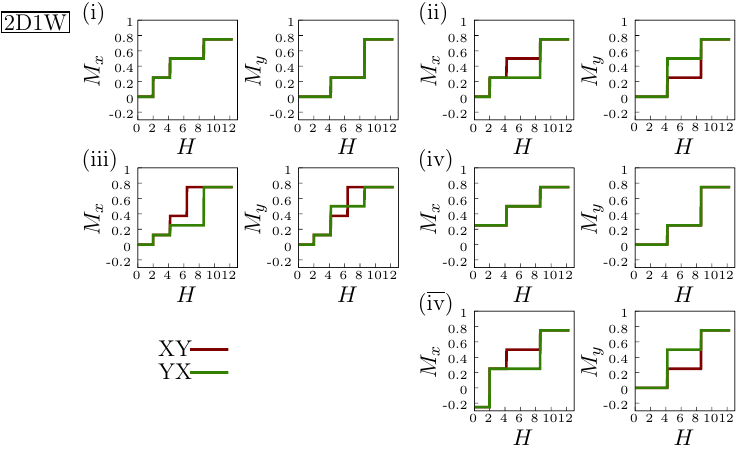}
\caption{Magnetization as function of the final external field $H$ for the bidirectional one way protocol with XY and YX paths, for $\alpha>2$ and $L=4$, and starting at the different degenerate states, defined in Fig.~\ref{Configl4}. Configurations (i)-(iii) are demagnetized, therefore their response is identical to that of their complementary configurations ($\overline{\textrm{i}}$)-($\overline{\textrm{iii}}$) with all spins flipped, whereas configuration (iv) has finite magnetization; it does not exhibit sequence dependence and the response of its complementary configuration ($\overline{\textrm{iv}}$) differs for XY vs YX.}
\label{Mag1wayA2p1}
\end{figure}

The impact of the sequence becomes evident when the magnetic field is strong enough to switch a vertex from Type~$I_3$ to Type~$II_3$ (i.e., when $H \ge 2(\alpha - 1)$). Depending on the arrangement of spins in the initial state, a specific path can cause different sets of spins on the lattice to flip at a given field. As a result, these spin flips can lead to different minimum energy states. For \( \alpha > 2 \), the magnetization from the 24 degenerate configurations reduce to five distinct groups, as illustrated in Fig.~\ref{Mag1wayA2p1}. The response is deterministic and among these five groups, three exhibit a path-dependent response.

In Fig.~\ref{DFGS02}, the dynamical changes in the configurations at the end of the protocol are shown for two different paths corresponding to configuration (iii.a) with \( \alpha > 2 \). The first excitation occurs at an external field of \( H = 1 \). In both the XY and YX paths, we observe that two Type \( II_2 \) vertices are present, leaving the configurations of the three-coordinated vertices unchanged.

\begin{figure}[t]
\centering   
\includegraphics[width=\columnwidth]{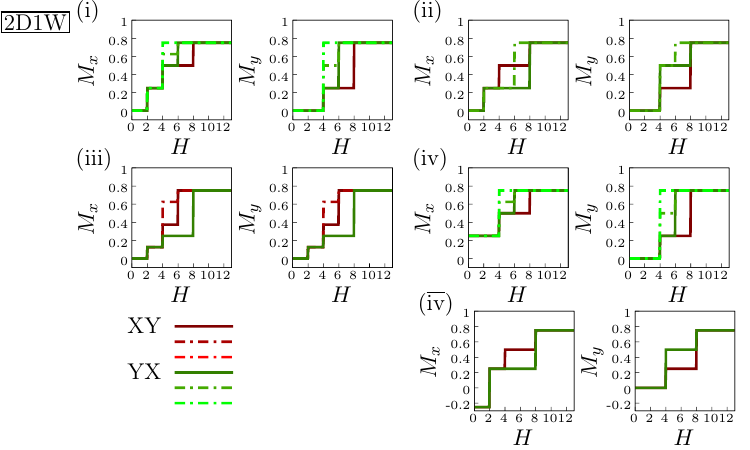}
\caption{Magnetization as function of the final external field $H$ for the bidirectional one way protocol for $\alpha\le2$ and $L=4$, starting from the different degenerate states, defined in Fig.~\ref{Configl4}. The different line styles correspond to the stochastic result obtained during the protocol.}
\label{Mag1wayA2}
\end{figure}

\begin{figure}[!h]
\centering   
\includegraphics[width=0.95\columnwidth]{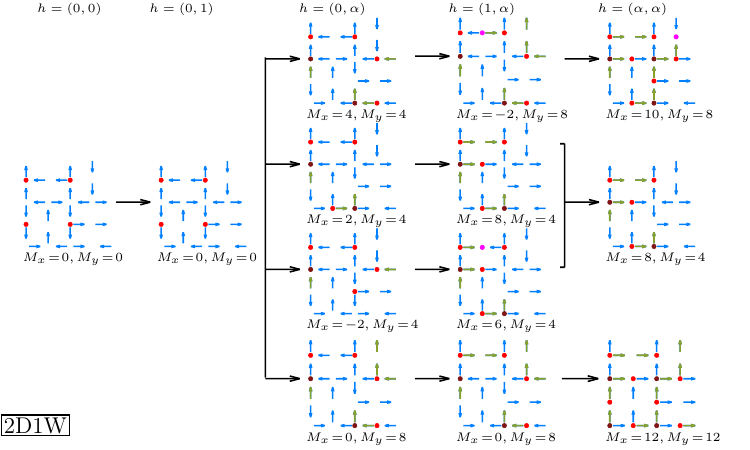}
\caption{The configurational changes during the bidirectional one way protocol along the XY path and starting from configuration (i.a) with $\alpha=2$. At $\vec{h} = (0,\alpha)$, we observe four intermediate possible minimum energy configurations. These four intermediate configurations reduce to three when the field in both directions reaches its final value, $\vec{h} = (\alpha,\alpha)$.}
\label{DFGS01}
\end{figure}

Furthermore, for any value of \(\alpha \leq 2\) when we measure the magnetization at the end of the protocol, we observe several possible values for the same external magnetic field, as seen in Fig.~\ref{Mag1wayA2}. The multiple outcomes occur due to bifurcations in the dynamics following spins randomly flipping in different sequences. This leads to multiple paths, resulting in different magnetization outcomes. This branching for one of the degenerate configurations (i.a) is shown in Fig.~\ref{DFGS01} for an external field of $H = \alpha$.

\begin{figure}[t]
\centering
\includegraphics[width=1\columnwidth]{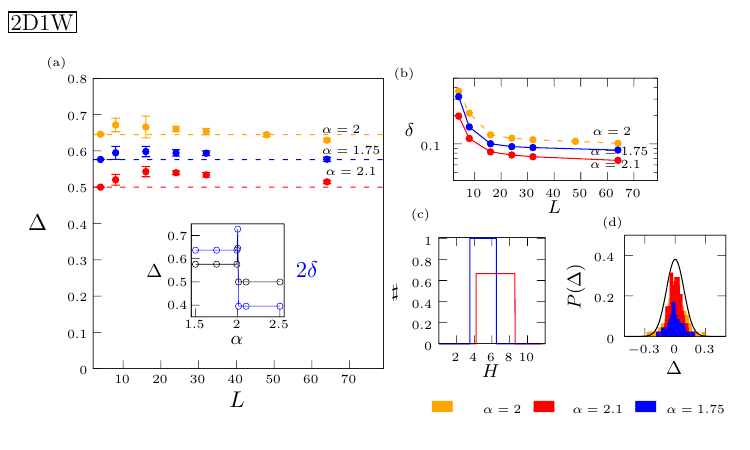}
\caption{a) The difference in magnetization between the XY and YX bidirectional one way protocols applied to the degenerate state for increasing system sizes. The dashed lines represent exact values obtained for $L=4$ for the different values of $\alpha$ indicated in the figure. Inset shows the behavior of $\Delta$ as function of $\alpha$ for $L=4$. b) Squared difference in magnetization vs. system size for different values of $\alpha$. c) Fraction of configurations that show sequence dependence as function of external magnetic field. d) The distribution of the difference ($\Delta$) for $L=24$. Black line is a Gaussian fit.}
\label{SystemsizeMag}
\end{figure}

We calculate the difference in magnetization at the end of XY and YX paths for the 2D1W protocol. Statistically, this difference is the same in both directions; hence, we use their average to define,
\be
\Delta=\frac{1}{2(\alpha-1)}\left[\int dh\left(M_x^{XY}-M_x^{YX}\right)+\int dh \left(M_y^{XY}-M_y^{YX}\right)\right] ,
\ee
where $\alpha - 1$ is a normalization constant, as the threshold external field is scaled by the factor $\alpha$. For the supercell with \( L=4 \), we carefully track the flipping of each spin to calculate the difference in average magnetization obtained during the XY and YX protocols for each initial configuration. As shown in Fig.~\ref{SystemsizeMag}a, we find that for \( \alpha > 2 \), the average value is \( \Delta = 0.5 \). For $\alpha\le 2$, the average difference in magnetization is larger than for $\alpha>2$ because of the stochastic value of magnetization observed at the end of the protocol. At \(\alpha = 2\), there is a singularity in the difference in magnetization. Furthermore, the fluctuations of this difference around the mean for \(\alpha = 2\) are more pronounced compared to other $\alpha$ values, further emphasizing the critical behavior at this value of $\alpha$. Additionally, we numerically calculate this difference using Monte Carlo simulations for larger systems, as exact analysis is challenging due to the permutations of spin flips and the large number of degenerate configurations. Interestingly, the average difference in magnetization remains finite and is close to the value obtained for \(L = 4\). 

Further, we calculate 
\be
\delta=\frac{1}{2(\alpha-1)}\left[\int dh\left(M_x^{XY}-M_x^{YX}\right)^2+\int dh \left(M_y^{XY}-M_y^{YX}\right)^2\right] ,
\ee
which, as shown in Fig.~\ref{SystemsizeMag}b decays with system size. Squaring the difference ensures that all values are non-negative. The multiple supercells comprising large systems might cancel the differences in some of the configurations. Thus, we observe saturation of the squared difference $\delta$ with increasing system size. In Fig.~\ref{SystemsizeMag}c, we also show the total number of degenerate configurations that display sequence-dependent responses at the end of the 2D1W protocol. For \(\alpha \leq 2\), all degenerate states exhibit a sequence-dependent response during an intermediate field; however, for \(\alpha > 2\), such response is exhibited only by two thirds of the configurations.

\begin{figure}[t]
\centering
\begin{subfigure}[b]{0.9\columnwidth}      
\includegraphics[width=\columnwidth]{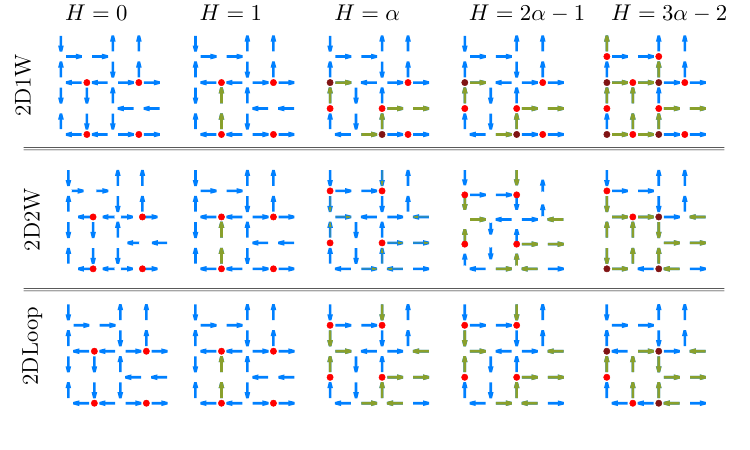}
\caption{Configuration (ii.b)}
\label{DCIIb}
\end{subfigure}
\begin{subfigure}[b]{0.9\columnwidth}
\includegraphics[width=\columnwidth]{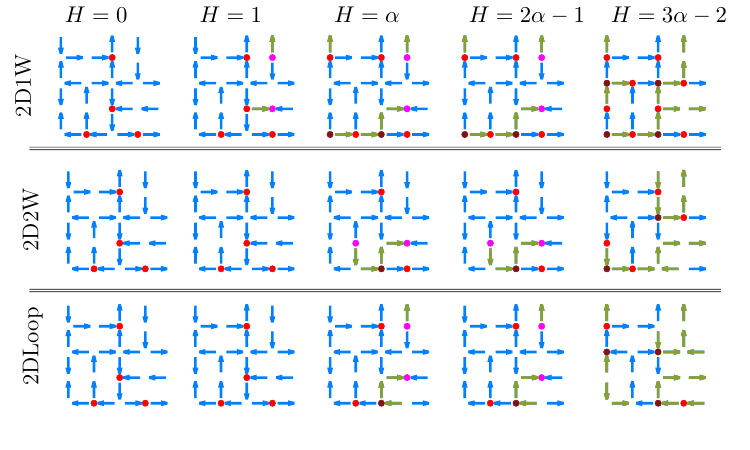}
\caption{Configuration (iii.a)} 
\label{DCIIIa}
\end{subfigure}
\caption{Configurational changes at the end of the different bidirectional protocols for $\alpha>2$. The final external field $H$ is listed above each column. a) For both 2D2W and 2Dloop protocols, configuration (ii.b) returns to a degenerate state with larger magnetization. b) For configuration (iii.a) with the 2D2W and 2DLoop protocols, the final configurations remain excited.}
\label{ConfigDynA2p1}
\end{figure}

\paragraph{Antiferromagnet state:}  
The antiferromagnet state in the Shakti lattice exhibits behavior similar to that in the square lattice when the external magnetic field is bidirectional and follows a one-way protocol. At a field strength $H > \alpha + 1$, all spins suddenly flip in the direction of the field for both XY and YX paths. Therefore, the antiferromagnet state does not display any sequence dependence and shows deterministic dynamics.

In the following two protocols, we will examine the memory of the Shakti lattice to determine whether the system retains its initial state or deviates from it when returning to zero field after the maximal external magnetic field is reached. This analysis will help us understand the effect of path dependence on memory retention. We can further illustrate this concept by examining hysteresis loops. The magnetization saturates at its limits for extreme positive and negative magnetic field values. If the external field is reversed before reaching these extremes, the system typically traces out partial loops that are connected to the main loop. Eventually, when the system returns to the same state within the main loop, this phenomenon is called return-point memory. The result can be characterized with various outcomes depending on the range of the external field. At low external field, the system ends in the degenerate state, at intermediate external field, there can be coexistence between the degenerate state and excited states, or there can be fluctuations among different excited states. 

\subsection{Two Way Protocol, 2D2W}

\begin{figure}[h!]
\centering\includegraphics[width=0.9\columnwidth]{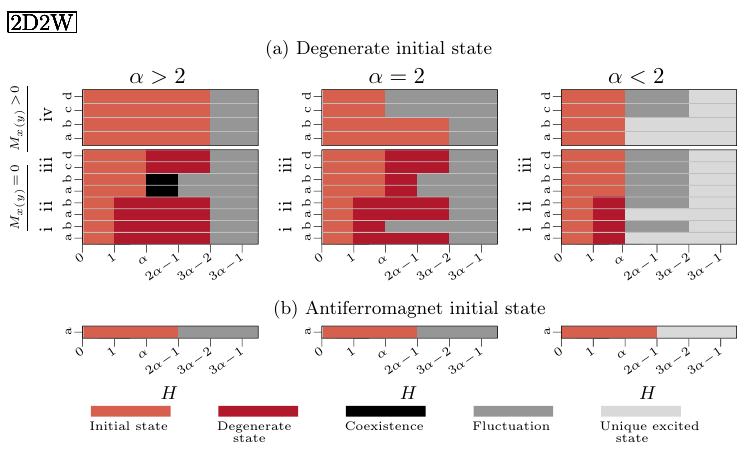}
\caption{Memory retention at the end of the bidirectional two-way protocol 2D2W as function of the maximal external field $H$ for different ranges of $\alpha$, starting at all degenerate states (a) or at the antiferromagnetic state (b). Colors indicate whether the system ends in its initial state (orange) or in some other degenerate state (red), may coexist between various possible excited or degenerate states (black), fluctuates between various excited states (dark gray), or ends in a unique excited state (light-gray).}
\label{BiD2w}
\end{figure}

\paragraph{Degenerate state:}
In the two way protocol, we return to zero external magnetic field by retracing the same path that was used to reach the maximal field, $H$, see Fig.~\ref{Prtime}b. We begin by analyzing the degenerate configurations for the Shakti supercell ($L=4$) using exact calculations for all states. After returning to zero field from varying maximal external fields, the system's final configurations are displayed in the middle row of Fig.~\ref{ConfigDynA2p1}. A field of \( H = 1 \) is the smallest value that can alter the energy landscape of the system. At this field, when a spin flip creates defects in a two-coordinated vertex (transitioning from Type \( I_2 \) to \( II_2 \)) at the end of the one-way protocol, the configuration returns to its initial degenerate configuration at the end of the two-way protocol. In contrast, at this minimal external field, if the spin that flips is between two-(Type~$II_2$)- and three-(Type~$II_3$)-coordinated vertices, the system returns to a minimum energy configuration that corresponds to degenerate state (iv) which has finite magnetization (see Fig.~\ref{Configl4}). This configuration obtained after returning to zero field is stable, and the outcome will differ from the configuration at which the process began.

In the presence of intermediate external magnetic fields, as long as not all spins have aligned with the field direction, the degenerate states will tend to return to one of the degenerate configurations that have larger magnetization. However, for configurations (iii.a-d), the outcomes can be either reverting to the degenerate states or failing to return to the degenerate state (as indicated by black in Fig.~\ref{BiD2w}a). The differences in outcomes for configurations~(iii.a-d) compared to other configurations arise from variations in the arrangement of defects; In configuration~(iii), the defects are concentrated along an axis, and the two-coordinated vertices (Type~$I_1$) are positioned in between the defect (Type~$II_3$) and non-defect (Type~$I_3$) states of the three-coordinated vertices. This arrangement necessitates flipping spins involving Type~$I_4$ and Type~$I_3$, transforming the landscape in which configurations with four-coordinated vertex in the first excited state to a minimum energy state under certain spin flip permutations. At higher external field, the system ends up in one of the many higher-energy states indicates by dark gray in Fig.~\ref{BiD2w} .

Moreover, the initial change in the external magnetic field primarily influences the direction of the finite magnetization of the degenerate states upon returning. For example, if we choose the XYYX path, the $x$ magnetization of the final returned configuration should surpass the $y$ magnetization.

In conclusion, the returned final minimum energy state is influenced not only by the arrangement of spins but also by the sequence in which the external field is applied. The returned state for the XYYX path of the 2D2W protocol is presented in Fig.~\ref{BiD2w}a for \( L=4 \). For the YXXY path, the outcome shows dynamics similar to those of XYYX, with the final minimum-energy states exhibiting magnetization in the opposite direction, indicating sequence-dependent response. Additionally, the fluctuating configurations switch from (iii.a,b) to (iii.c,d) due to the rotational symmetry of these configurations.

We reexamined the return-point memory for \(\alpha=2\) and found outcomes that differ from those observed for \(\alpha>2\). For \(\alpha=2\), our analytical observations indicate a return to one of the degenerate states as long as the external field strength \(H\) is less than $3\alpha-2$, as shown in Fig.~\ref{BiD2w}a. However, most configurations display outcomes similar to those seen for \(\alpha>2\), with the exception of configuration~(iii), which exhibits stochastic behavior between the degenerate state and excited states. Additionally, for configurations~(i.a) and~(ii.b) early breakdown of return-point memory in the lower external field predominantly depends on the arrangement of spins. For $\alpha<2$, at a small external field, the system loses its ability to return to one of the degenerate states. For higher fields $(H > \alpha)$, some configurations return to unique excited states (as indicated by light gray in Fig.~\ref{BiD2w}) or fluctuate among different excited states.

\paragraph{Antiferromagnetic state:} The response of the antiferromagnetic state to increasing the field by the 2D1W protocol is similar to that of the square lattice, without any sequence dependence. However, for $\alpha \ge 2$, following the decrease of the magnetic field in the 2D2W protocol ends in sequence dependent response. For $\alpha<2$, for high enough external field the system reaches a state where all spins are aligned in the direction of the external field and will not able to return to any other states while returning to zero magnetic field, as shown in Fig.~\ref{BiD2w}b. Further, for $\alpha<2$, the system ends in a unique excited state at higher external field, hence the dynamics is deterministic. For $\alpha\ge 2$,  the response is stochastic and the system relaxes to one of its many excited states.

\begin{figure}[!h]
\centering
\includegraphics[width=0.9\columnwidth]{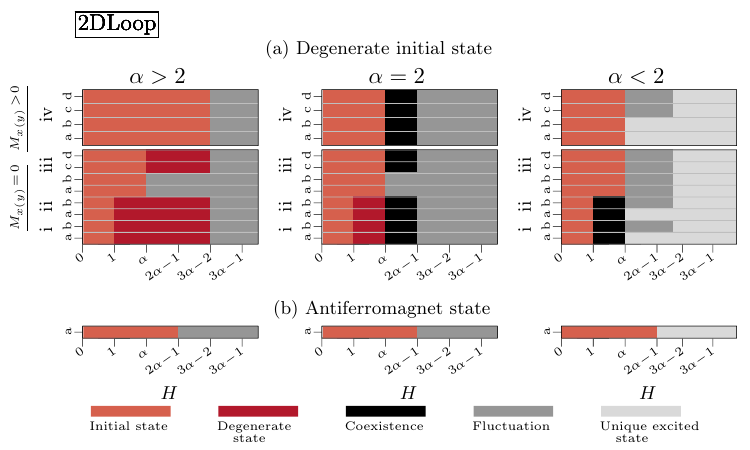}
\caption{Memory retention at the end of the bidirectional loop protocol 2DLoop as function of the maximal external field $H$ for different ranges of $\alpha$, starting at all degenerate states (a) or at the antiferromagnetic state (b). Colors indicate whether the system ends in its initial state (orange) or in some other degenerate state (red), may coexist between various possible excited or degenerate states (black), fluctuates between various excited states (dark gray), or ends in a unique excited state (light-gray).}
\label{BiDLoop}
\end{figure}

\subsection{Loop Protocol, 2DLoop}

\paragraph{Degenerate state:} In contrast to the 2D2W protocol, we modify the order of the return path, creating a loop that returns to zero magnetic field, as illustrated in Fig.~\ref{Prtime}c. The outcomes for the 2Loop and 2D2W protocols are identical for most configurations except fluctuating dynamics observed in configuration~(iii) for some range of the external magnetic field. In the loop protocol, the path taken to reach the final external field~\(H\) differs from the return path. As a result, the sequence of spin flips at any given field varies because flips of orthogonal spins are more prevalent during the return path. Therefore, for configurations~(iii.a-b), the specific sequence of spin flips that leads to the degenerate configurations is not attainable. As shown in Fig.~\ref{BiDLoop}a, we do not observe fluctuating outcomes, except for the singular case $\alpha=2$. Additionally, when comparing the configurational changes in the two-way and loop protocols (compare to Fig.~\ref{ConfigDynA2p1} above), we find that the return to the degenerate states at the external field $H \le 3\alpha-2$ for configuration~(i.a) is identical, and the spin flips are also the same. However, for configuration~(iii.a), we begin to see differences in the outcomes for $H >\alpha$.

In the loop protocol, we begin to observe fluctuating outcomes for intermediate external fields ($\alpha \leq H \leq 2\alpha-1$) when \(\alpha=2\) across most configurations (see Fig.~\ref{BiDLoop}a). For higher external magnetic fields ($H > 2\alpha-1$), the system loses its memory and ends in one of the higher-energy state, thus fails to return to the initial, degenerate state. The sequence-dependent response is observed for intermediate and higher external fields until all the spins are aligned in the direction of the field, and the end state is the same as the initial state.

Furthermore, the threshold external magnetic field is influenced by \(\alpha\); consequently, lower values of \(\alpha\) correspond to an earlier breakdown of memory retention under external fields. Our observations reveal that for \(\alpha<2\), the system does not return to any of the degenerate configurations beyond an external field $H = \alpha$. For \(\alpha < 2\), the memory to retain the degenerate state breaks down very early with the application of an external magnetic field as depicted in Fig~\ref{cutoff}c. 

\begin{figure}[h!]
\centering
\includegraphics[width=\columnwidth]{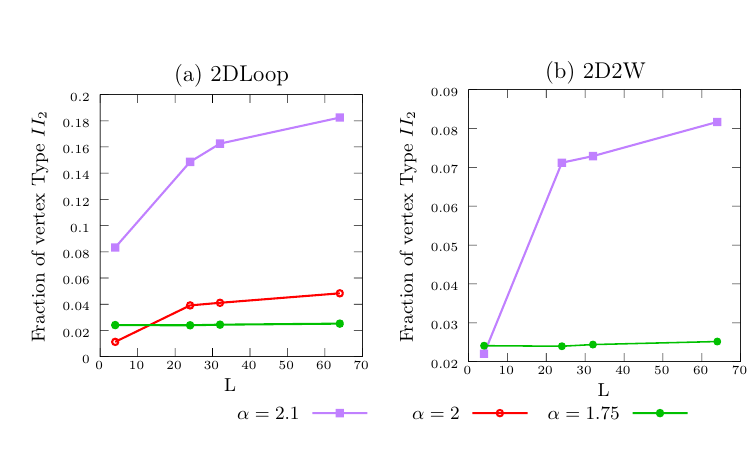}
\caption{System size dependence of the fraction of vertices of Type $II_2$ at the end of the 2D2Loop and 2D2W protocols, starting from the degenerate state for different values of $\alpha$ and with an external field of $H=\alpha$.}
\label{VertexA2}
\end{figure}

For large system sizes, we calculate the fraction of vertices at the end of various protocols under the influence of different external fields \( H \). Our results indicate that, unlike the minimal $L=4$ system, larger systems consistently exhibit a finite number of defects at intermediate external fields (\( H \ge \alpha \)). We specifically focus on Type~\( II_2 \) vertices that are generated during the process, as this helps deepen our understanding on the dependence of degenerate initial state on \( \alpha \) and the number of defects is larger in the 2DLoop protocol. This quantifies the deviation of the final configuration from the initial degenerate state hence affecting the return point memory.

For both the loop and two-way protocols, the number of Type~\( II_2 \) defects remains finite and independent of system size when \( \alpha < 2 \). However, at the critical point \( \alpha = 2 \), the 2D2W protocol does not produce any Type~\( II_2 \) defects, as shown in Fig.~\ref{VertexA2}b. For~\( \alpha > 2 \), the fraction of Type~$II_2$ defects increases monotonically with system size in both protocols. This behavior is attributed to the fact that the degenerate state is not the ground state, and the deviation from the ground state is much larger for \( \alpha > 2 \).

\paragraph{Antiferromagnetic state:} We investigate the effects of applying the two way and loop protocols to the antiferromagnetic configurations. Similar to the behavior observed in a square lattice, we notice a sudden flipping of all spins to align with the direction of the external magnetic field at a threshold value \(H_c\). This value of $H_c$ correspond to the field required to flip the spin between vertex Type~$I_3$ and~$II_2$ and hence is not the same as in the square lattice. Once all the spins are aligned with the field, the system loses its memory as it returns to zero magnetic field. This behavior holds true regardless of whether the return is achieved through the two-way or the loop protocol, as illustrated in Figs.~\ref{BiD2w}b and~\ref{BiDLoop}b.

\section{Discussion}
\label{sec:conclusions}

\begin{figure}[t]
\centering
\includegraphics[width=0.9\columnwidth]{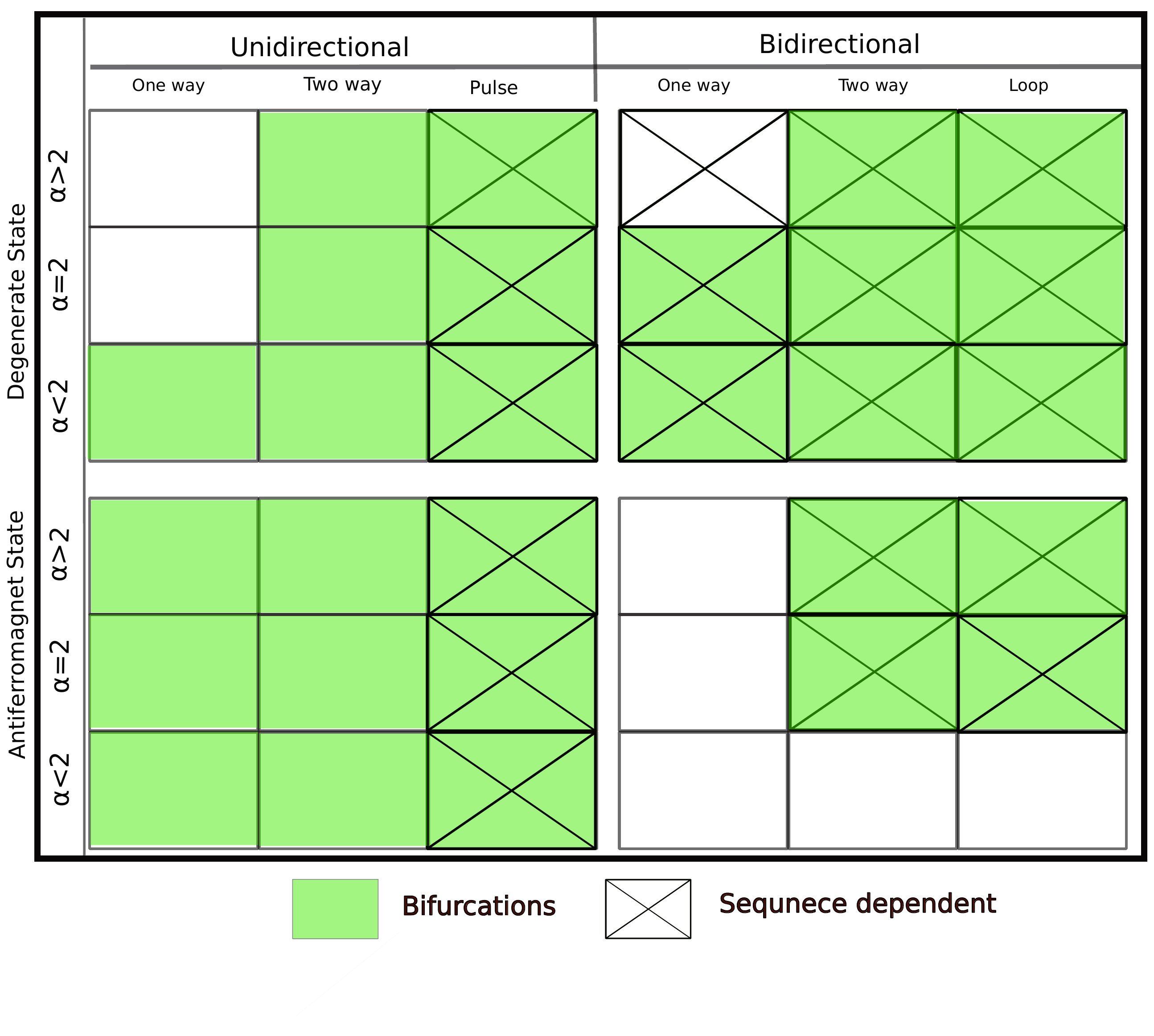}
\caption{Summary of results for degenerate and antiferromagnet states. Green boxes represent bifurcations, whereas crossed boxes indicate sequence dependence.}
\label{fig:table}
\end{figure}

We have systematically investigated memory by probing the response of square and Shakti artificial spin ice to various protocols for dynamically changing an in-plane external magnetic field. This response is demonstrated through changes in magnetization and positioning of excitations, which are fundamental to any artificial spin ice system. In square spin ice, uniformity in lattice structure with a single coordination number results in a response that does not depend on the sequence of changing the external field. For Shakti, the ground state switches from antiferromagnetic to a topologically-constrained disordered and thus degenerate state depending on the ratio $\alpha$ of parallel and perpendicular spins interaction strengths. Under the influence of an external magnetic field, both these states may display sequence-dependent response when the field is applied in a unidirectional pulse protocol or in bidirectional protocols, as the applied field must break symmetry either in time or space. This is interesting because a richness that is absent in the ordered square spin ice is instead present even in the ordered version of Shakti. The reason, we propose, lays in the fact that for ordered Shakti ($\alpha>2$), unlike in square ice, the topological state represent a degenerate manifold of excitations. We include a summary of our results in Fig.~\ref{fig:table}, where the sequence-dependence response is marked with a cross for the given parameter, and green-colored boxes indicate bifurcations that lead to coexistence or fluctuations. 

For the 2D1W protocol we observe two different behavior, depending on the energetics: in the $\alpha>2$ case---which is the Shakti spin ice whose ground state is ordered---we observe deterministic dynamics, with an outcome that depends on the initial configuration. Instead, for $\alpha\le2$---which is the Shakti spin ice whose ground state is disordered but topologically ordered---the outcomes fluctuate, resulting in several possible outcomes at intermediate fields. Importantly, if in the degenerate Shakti ($\alpha<2$) we start from its own proper degenerate ground state (Fig.~\ref{lattice}c, left), we see that all protocols can lead to {\it stochastic response}. Instead, if we prepare the degenerate Shakti in the ordered antiferromagnetic ground state (Fig.~\ref{lattice}c, right), we obtain a stochastic response only for unidirectional driving protocols. Notably, when $\alpha=2$, in the degenerate state, the response peaks. Furthermore in the 2D1W protocol applied on the antiferromagnetic state, there is no sequence dependence. 

Additionally, we studied whether the system retains its memory to return to the initial state for two protocols: the 2D2W protocol and the 2DLoop protocol. We found that in the 2DLoop protocol, the intertwined flipping of \(x\) and \(y\) spins can often destroy the memory of the path. In contrast, the 2D2W protocol allows the system to retrieve the initial states for large maximal fields. 

Our study highlights the crucial role of the interaction strength ratio \(\alpha\) in determining the impact of the external fields. For a disordered Shakti \(\alpha < 2\) the behavior of the minimal supercell under both the two-way and loop protocols is similar. However, for \(\alpha \geq 2\), the two-way protocol retains its memory until all the spins are aligned with the external field and vice versa for the antiferromagnet state. This behavior of the minimal supercell can be averaged and mapped to understand the behavior of larger systems. 

Future work might investigate why topologically protected disorder can lead to the reported memory and stochasticity. Shakti's ground state can be mapped into a dimer model, but its elementary kinetics, such the parallel dimer flips, are not direct but go through the creation of topological charge pairs that might or might not recombine (see Supplementary information of Ref.~\cite{lao2018classical}) perhaps providing a mechanism for the observed stochasticity. Clearly, memory rests not only on phases but also, indeed mostly, on kinetics, and therefore it should be interesting to investigate how system of similar phases and different kinetics perform under probes of memory. For instance, the ground state of Santa Fe spin ice~\cite{zhang2021string} can also be mapped into a dimer model, but with an entirely different kinetics, as the role of dimers is played by strings that can wiggle, elongate, or contract~\cite{zhang2023topological}. Kagome spin ice might also be a good candidate for protocols such as those described here, to discern between the effect of disorder and its topological protection. There, the Ice~I phase is disordered but not topological, whereas the Ice~II phase is~\cite{chern2011two, moller2009magnetic}. Finally, it would be interesting to explore the memory of sheared square lattice, in which the energy spectrum of vertex types is modified such that excitations are topologically ordered~\cite{oguz2020c}.

In summary, we have shown that topologically constrained disorder can act as a powerful, if hitherto under-appreciated, carrier of memory. Whereas the antiferromagnetically ordered square artificial spin ice responds predictably and without history dependence, Shakti's geometry, because its degenerate ground state manifold, exhibits rich protocol-dependent dynamics ranging from fully deterministic return to the initial ground state to pronounced path-dependent bifurcations and stochastic final configurations, revealing a clear hierarchy of memory sensitivity governed by small changes of the involved parameters. Our results propose a wealth of new phenomenology largely unexpected in disordered systems, and which future works will have to understand. In the broader context, it demonstrates that topological protection, already known to limit ergodicity~\cite{henley2011classical} can be employed for exquisitely sensitive, programmable responses.

\section*{Acknowledgments} We thank Chaviva Sirote-Katz for helpful discussions. The work of C.N. was performed by under the auspicies of the U.S. Department of Energy through the Los Alamos National Laboratory, operated by Triad National Security, LLC, for the National Nuclear Security Administration of U.S. Department of Energy (Contract No. 89233218CNA000001). This research was supported in part by the Israel Science Foundation Grant No. 1899/20. Y.S. thanks the Center for Nonlinear Studies at Los Alamos National Laboratory for its hospitality.

\section*{References}
\bibliography{shaktiR}

@article{branford2012emerging,
  title={Emerging chirality in artificial spin ice},
  author={Branford, WR and Ladak, Sam and Read, Daniel E and Zeissler, K and Cohen, LF},
  journal={Science},
  volume={335},
  pages={1597--1600},
  year={2012},
  url = {https://doi.org/10.1126/science.1211379}
}

@article{caravelli2022artificial,
  title={Artificial spin ice phase-change memory resistors},
  author={Caravelli, Francesco and Chern, Gia-Wei and Nisoli, Cristiano},
  journal={New J. Phys.},
  volume={24},
  pages={023020},
  year={2022},
  url = {https://doi.org/10.1088/1367-2630/ac4c0a}
}

@article{gilbert2016emergent,
  title={Emergent reduced dimensionality by vertex frustration in artificial spin ice},
  author={Gilbert, Ian and Lao, Yuyang and Carrasquillo, Isaac and O’Brien, Liam and Watts, Justin D and Manno, Michael and Leighton, Chris and Scholl, Andreas and Nisoli, Cristiano and Schiffer, Peter},
  journal={Nat. Phys.},
  volume={12},
  pages={162--165},
  year={2016},
  url = {https://doi.org/10.1038/nphys3520}
}

@article{chern2017magnetotransport,
  title={Magnetotransport in artificial kagome spin ice},
  author={Chern, Gia-Wei},
  journal={Phys. Rev. Appl.},
  volume={8},
  pages={064006},
  year={2017},
  url = {https://doi.org/10.1103/PhysRevApplied.8.064006}
}

@article{le2017understanding,
  title={Understanding magnetotransport signatures in networks of connected permalloy nanowires},
  author={Le, BL and Park, Jungsik and Sklenar, Joseph and Chern, G-W and Nisoli, Cristiano and Watts, JD and Manno, M. and Rench, D. W. and Samarth, N. and Leighton, C. and  Schiffer, P.},
  journal={Phys. Rev. B},
  volume={95},
  pages={060405},
  year={2017},
  url = {https://doi.org/10.1103/PhysRevB.95.060405}
}

@article{schanilec2022approaching,
  title={Approaching the topological low-energy physics of the F model in a two-dimensional magnetic lattice},
  author={Sch{\'a}nilec, V and Brunn, O and Hor{\'a}{\v{c}}ek, M and Kr{\'a}tk{\`y}, S and Meluz{\'\i}n, P and {\v{S}}ikola, T and Canals, Benjamin and Rougemaille, N},
  journal={Phys. Rev. Lett.},
  volume={129},
  pages={027202},
  year={2022},
  url = {https://doi.org/10.1103/PhysRevLett.129.027202}
}

@article{perrin2016extensive,
  title={Extensive degeneracy, Coulomb phase and magnetic monopoles in artificial square ice},
  author={Perrin, Yann and Canals, Benjamin and Rougemaille, Nicolas},
  journal={Nature},
  volume={540},
  pages={410--413},
  year={2016},
  url = {https://doi.org/10.1038/nature20155}
}

@article{moller2009magnetic,
  title={Magnetic multipole analysis of kagome and artificial spin-ice dipolar arrays},
  author={M{\"o}ller, Gunnar and Moessner, Roderich},
  journal={Phys. Rev. B},
  volume={80},
  pages={140409},
  year={2009},
  url = {https://doi.org/10.1103/PhysRevB.80.140409}
}

@article{porro2013exploring,
  title={Exploring thermally induced states in square artificial spin-ice arrays},
  author={Porro, JM and Bedoya-Pinto, Amilcar and Berger, Andreas and Vavassori, Paolo},
  journal={New J. Phys.},
  volume={15},
  pages={055012},
  year={2013},
  url = {https://doi.org/10.1088/1367-2630/15/5/055012}
}

@article{libal2012hysteresis,
  title={Hysteresis and return-point memory in colloidal artificial spin ice systems},
  author={Lib{\'a}l, A and Reichhardt, C and Olson Reichhardt, CJ},
  journal={Phys. Rev. E},
  volume={86},
  pages={021406},
  year={2012},
  url = {https://doi.org/10.1103/PhysRevE.86.021406}
}

@article{chern2011two,
  title={Two-stage ordering of spins in dipolar spin ice on the kagome lattice},
  author={Chern, Gia-Wei and Mellado, Paula and Tchernyshyov, O},
  journal={Phys. Rev. Lett.},
  volume={106},
  pages={207202},
  year={2011},
  url = {https://doi.org/10.1103/PhysRevLett.106.207202}
}

@article{wysin2013dynamics,
  title={Dynamics and hysteresis in square lattice artificial spin ice},
  author={Wysin, Gary M and Moura-Melo, WA and M{\'o}l, LAS and Pereira, AR},
  journal={New J. Phys.},
  volume={15},
  pages={045029},
  year={2013},
  url = {https://doi.org/10.1088/1367-2630/15/4/045029}
}

@article{Chern2013,
  title = {Degeneracy and Criticality from Emergent Frustration in Artificial Spin Ice},
  author = {Chern, G-W and Morrison, M. J. and Nisoli, C},
  journal = {Phys. Rev. Lett.},
  volume = {111},
  pages = {177201},
  year = {2013},
  url = {https://doi.org/10.1103/PhysRevLett.111.177201}
}

@article{lieb1967residual,
  title={Residual entropy of square ice},
  author={Lieb, Elliott H},
  journal={Phys. Rev.},
  volume={162},
  pages={162},
  year={1967},
  url = {https://doi.org/10.1103/PhysRev.162.162}
}

@article{nisoli2010effective,
  title={Effective Temperature in an Interacting Vertex System: {Theory} and Experiment on Artificial Spin Ice},
  author={Nisoli, Cristiano and Li, Jie and Ke, Xianglin and Garand, D and Schiffer, Peter and Crespi, Vincent H},
  journal={Phys. Rev. Lett.},
  volume={105},
  pages={047205},
  year={2010},
  url = {https://doi.org/10.1103/PhysRevLett.105.047205}
}

@article{nisoli2020topological,
  title={Topological order of the {Rys} {F}-model and its breakdown in realistic square spin ice: {Topological} sectors of {Faraday} loops},
  author={Nisoli, Cristiano},
  journal={Europhys. Lett.},
  volume={132},
  pages={47005},
  year={2020},
  url = {https://doi.org/10.1209/0295-5075/132/47005}
}

@article{henley2010coulomb,
  title={The “Coulomb phase” in frustrated systems},
  author={Henley, Christopher L},
  journal={Annu. Rev. Condens. Matter Phys.},
  volume={1},
  pages={179--210},
  year={2010},
  url = {https://doi.org/10.1146/annurev-conmatphys-070909-104138}
}

@article{henley2011classical,
  title={Classical height models with topological order},
  author={Henley, Christopher L},
  journal={J Phys: Condens. Matter},
  volume={23},
  pages={164212},
  year={2011},
  url = {https://doi.org/10.1088/0953-8984/23/16/164212}
}

@article{regev2015reversibility,
  title={Reversibility and criticality in amorphous solids},
  author={Regev, Ido and Weber, John and Reichhardt, Charles and Dahmen, Karin A and Lookman, Turab},
  journal={Nat. Commun.},
  volume={6},
  pages={8805},
  year={2015},
  url = {https://doi.org/10.1038/ncomms9805}
}

@article{ritort2003glassy,
  title={Glassy dynamics of kinetically constrained models},
  author={Ritort, Felix and Sollich, Peter},
  journal={Adv. Phys.},
  volume={52},
  pages={219--342},
  year={2003},
  url = {https://doi.org/10.1080/0001873031000093582}
}

@article{merrigan2021topologically,
  title={Topologically protected steady cycles in an icelike mechanical metamaterial},
  author={Merrigan, Carl and Nisoli, Cristiano and Shokef, Yair},
  journal={Phys. Rev. Res.},
  volume={3},
  pages={023174},
  year={2021},
  url = {https://doi.org/10.1103/PhysRevResearch.3.023174}
}

@article{yasuda2021mechanical,
  title={Mechanical computing},
  author={Yasuda, Hiromi and Buskohl, Philip R and Gillman, Andrew and Murphey, Todd D and Stepney, Susan and Vaia, Richard A and Raney, Jordan R},
  journal={Nature},
  volume={598},
  pages={39--48},
  year={2021},
  url = {https://doi.org/10.1038/s41586-021-03623-y}
}

@article{stamps2014unhappy,
  title={The unhappy wanderer},
  author={Stamps, Robert L},
  journal={Nat. Phys.},
  volume={10},
  pages={623--624},
  year={2014},
  url = {https://doi.org/10.1038/nphys3072}
}

@article{lao2018classical,
  title={Classical topological order in the kinetics of artificial spin ice},
  author={Lao, Yuyang and Caravelli, Francesco and Sheikh, Mohammed and Sklenar, Joseph and Gardeazabal, Daniel and Watts, Justin D and Albrecht, Alan M. and Scholl, Andreas and Dahmen, Karin and Nisoli, Cristiano and Schiffer, Peter},
  journal={Nat. Phys.},
  volume={14},
  pages={723--727},
  year={2018},
  url = {https://doi.org/10.1038/s41567-018-0077-0}
}

@article{gilbert2014emergent,
  title={Emergent ice rule and magnetic charge screening from vertex frustration in artificial spin ice},
  author={Gilbert, Ian and Chern, Gia-Wei and Zhang, Sheng and O’Brien, Liam and Fore, Bryce and Nisoli, Cristiano and Schiffer, Peter},
  journal={Nat. Phys.},
  volume={10},
  pages={670--675},
  year={2014},
  url = {https://doi.org/10.1038/nphys3037}
}

@article{Carolina2023,
  title = {Geometrical control of topological charge transfer in Shakti-Cairo colloidal ice},
  author = {Rodríguez-Gallo, C and Ortiz-Ambriz, A and Nisoli, C and Tierno, P},
  journal = {Comm. Phys.},
  volume = {6},
  pages = {113},
  year = {2023},
  url = {https://doi.org/10.1038/s42005-023-01236-7}
}

@article{Ding2022,
    author = {Ding, J and van Hecke, M},
    title = {Sequential snapping and pathways in a mechanical metamaterial},
    journal = {J. Chem. Phys.},
    volume = {156},
    pages = {204902},
    year = {2022},
    url = {https://doi.org/10.1063/5.0087863}
}

@article{Gilbert2016,
    author = {Gilbert, I and Nisoli, C and Schiffer, P},
    title = {Frustration by design},
    journal = {Phys. Today},
    volume = {69},
    pages = {54-59},
    year = {2016},
   url = {https://doi.org/10.1063/PT.3.3237}
}

@article{Kapaklis2012,
author = {Kapaklis, Vassilios and Arnalds, Unnar B and {Harman-Clarke}, Adam and Papaioannou, Evangelos Th and Karimipour, Masoud and Korelis, Panagiotis and Taroni, Andrea and Holdsworth, Peter C W and Bramwell, Steven T and Hj{\"o}rvarsson, Bj{\"o}rgvin},
title = {Melting artificial spin ice},
volume = {14},
pages = {035009},
journal = {New J. Phys.},
year = {2012},
url = {https://doi.org/10.1088/1367-2630/14/3/035009}
}

@article{Lieb1967,
  title = {Exact Solution of the {$F$} Model of An Antiferroelectric},
  author = {Lieb, E. H.},
  journal = {Phys. Rev. Lett.},
  volume = {18},
  pages = {1046 - 1048},
  year = {1967},
  url = {https://doi.org/10.1103/PhysRevLett.18.1046}
}

@article{Ladak2010,
title={Direct observation of magnetic monopole defects in an artificial spin-ice system},
author = {Ladak, S. and Read, D. E. and Perkins, G. K. and  Cohen, L. F. and Branford, W. R.},
journal={Nat. Phys.},
volume={6},
pages={359-363},
year={2010},
url = {https://doi.org/10.1038/nphys1628}
}

@article{Liu2024,
author = {Liu, J.  and Teunisse, M. and Korovin, G.  and Vermaire I. R. and Jin, L.  and Bense, H.  and Hecke, M. },
title = {Controlled pathways and sequential information processing in serially coupled mechanical hysterons},
journal = {Proc. Natl. Acad. Sci.},
volume = {121},
pages = {e2308414121},
year = {2024},
url = {https://doi.org/10.1073/pnas.2308414121}
}

@article{Lindeman2025,
title={Multi-dimensional memory in low-friction granular materials},
author={Lindeman, C M},
year={2025},
journal={Soft Matter},
volume={21},
pages={4890-4897},
url = {https://doi.org/10.1039/D5SM00470E }
}

@article{Morrison2013,
author = {Morrison, M. .J and Nelson, T. R. and Nisoli, C.},
title = {Unhappy vertices in artificial spin ice: new degeneracies from vertex frustration},
journal = {New J. Phys.},
year = {2013},
volume = {15},
pages = {045009},
url = {https://doi.org/10.1088/1367-2630/15/4/045009}
}

@article{Mengotti2011,
title={Real-space observation of emergent magnetic monopoles and associated Dirac strings in artificial kagome spin ice},
author={ Mengotti, E. and Heyderman, L. J. and Rodríguez, A F and Nolting, F and Hügli, R. V. and Braun, H-B},
year={2011},
journal ={Nat. Phys.},
volume={7},
pages={68-74},
url = {https://doi.org/10.1038/nphys1794}
}

@article{zhang2021string,
  title={String phase in an artificial spin ice},
  author={Zhang, Xiaoyu and Duzgun, Ayhan and Lao, Yuyang and Subzwari, Shayaan and Bingham, Nicholas S and Sklenar, Joseph and Saglam, Hilal and Ramberger, Justin and Batley, Joseph T and Watts, Justin D and Bromley, Daniel and Chopdekar, Rajesh V. and O’Brien, Liam and Leighton, Chris and Nisoli, Cristiano and Schiffer, Peter},
  journal={Nat. Commun.},
  volume={12},
  pages={6514},
  year={2021},
  url = {https://doi.org/10.1038/s41467-021-26734-6}
}

@article{Meulblok2025,
title={Path-dependency and emergent computing under vectorial driving},
author={Meulblok, C M and Singh, A and Labousse, M and Hecke, M Van},
year={2025},
journal={arXiv:2503.07764},
url= {https://doi.org/10.48550/arXiv.2503.07764}
}

@article{keim2021multiperiodic,
  title={Multiperiodic orbits from interacting soft spots in cyclically sheared amorphous solids},
  author={Keim, Nathan C and Paulsen, Joseph D},
  journal={Sci. Adv.},
  volume={7},
  pages={eabg7685},
  year={2021},
  url = {https://doi.org/10.1126/sciadv.abg7685}
}

@article{Nisoli2013,
  title = {Colloquium: Artificial spin ice: Designing and imaging magnetic frustration},
  author = {Nisoli, C. and Moessner, R. and Schiffer, P.},
  journal = {Rev. Mod. Phys.},
  volume = {85},
  pages = {1473-1490},
  year = {2013},
  url = {https://doi.org/10.1103/RevModPhys.85.1473}
}

@article{sendetskyi2019continuous,
  title={Continuous magnetic phase transition in artificial square ice},
  author={Sendetskyi, Oles and Scagnoli, Valerio and Leo, Na{\"e}mi and Anghinolfi, Luca and Alberca, Aurora and L{\"u}ning, Jan and Staub, Urs and Derlet, Peter Michael and Heyderman, Laura Jane},
  journal={Phys. Rev. B},
  volume={99},
  pages={214430},
  year={2019},
  url = {https://doi.org/10.1103/PhysRevB.99.214430}
}

@article{morgan2011thermal,
  title={Thermal ground-state ordering and elementary excitations in artificial magnetic square ice},
  author={Morgan, Jason P and Stein, Aaron and Langridge, Sean and Marrows, Christopher H},
  journal={Nat. Phys.},
  volume={7},
  pages={75--79},
  year={2011},
  url = {https://doi.org/10.1038/nphys1853}
}

@article{zhang2013crystallites,
  title={Crystallites of magnetic charges in artificial spin ice},
  author={Zhang, Sheng and Gilbert, Ian and Nisoli, Cristiano and Chern, Gia-Wei and Erickson, Michael J and O’brien, Liam and Leighton, Chris and Lammert, Paul E and Crespi, Vincent H and Schiffer, Peter},
  journal={Nature},
  volume={500},
  pages={553--557},
  year={2013},
  url = {https://doi.org/10.1038/nature12399}
}

@article{Nisoli2007,
  title = {Ground State Lost but Degeneracy Found: The Effective Thermodynamics of Artificial Spin Ice},
  author = {Nisoli, C. and Wang, R. and Li, J. and McConville, W. F. and Lammert, P. E. and Schiffer, P. and Crespi, V. H.},
  journal = {Phys. Rev. Lett.},
  volume = {98},
  pages = {217203},
  year = {2007},
  url = {https://doi.org/10.1103/PhysRevLett.98.217203}
}

@article{Sirote2024,
  title = {Emergent disorder and mechanical memory in periodic metamaterials},
  author = {Sirote-Katz, C. and Shohat, D. and Merrigan, C. and Lahini, Y. and Nisoli, C. and Shokef, Y.},
  journal = {Nat. Commun.},
  volume = {15},
  pages = {4008},
  year = {2024},
  url = {https://doi.org/10.1038/s41467-024-47780-w}
}

@article{kudasov2006steplike,
  title={Steplike magnetization in a spin-chain system: Ca3 Co2 O6},
  author={Kudasov, Yuri B},
  journal={Phys. Rev. Lett.},
  volume={96},
  pages={027212},
  year={2006},
  url = {https://doi.org/10.1103/PhysRevLett.96.027212}
}

@article{Sandra2020,
  title = {Advances in artificial spin ice},
  author = {Skj{\ae}rv{\o}, Sandra H. and  Marrows, Christopher H. and Stamps, Robert L. and Heyderman, Laura J.},
  journal = {Nat. Rev. Phys.},
  volume = {2},
  pages = {13-28},
  year = {2020},
  url = {https://doi.org/10.1038/s42254-019-0118-3}
}

@article{zhang2023topological,
  title={Topological kinetic crossover in a nanomagnet array},
  author={Zhang, Xiaoyu and Fitez, Grant and Subzwari, Shayaan and Bingham, Nicholas S and Chioar, Ioan-Augustin and Saglam, Hilal and Ramberger, Justin and Leighton, Chris and Nisoli, Cristiano and Schiffer, Peter},
  journal={Science},
  volume={380},
  pages={526--531},
  year={2023},
  url = {https://doi.org/10.1126/science.add6575}
}

@article{Shohat2022,
author = {Shohat, D.  and Hexner, D.  and Lahini, Y. },
title = {Memory from coupled instabilities in unfolded crumpled sheets},
journal = {Proc. Natl. Acad. Sci.},
volume = {119},
pages = {e2200028119},
year = {2022},
url = {https://doi.org/10.1073/pnas.2200028119}
}

@article{gilbert2015direct,
  title={Direct visualization of memory effects in artificial spin ice},
  author={Gilbert, Ian and Chern, Gia-Wei and Fore, Bryce and Lao, Yuyang and Zhang, Sheng and Nisoli, Cristiano and Schiffer, Peter},
  journal={Phys. Rev. B},
  volume={92},
  pages={104417},
  year={2015},
  url = {https://doi.org/10.1103/PhysRevB.92.104417}
}

@article{nisoli2017deliberate,
  title={Deliberate exotic magnetism via frustration and topology},
  author={Nisoli, Cristiano and Kapaklis, Vassilios and Schiffer, Peter},
  journal={Nat. Phys.},
  volume={13},
  pages={200--203},
  year={2017},
  url = {https://doi.org/10.1038/nphys4059}
}

@article{Stopfel2018,
  title = {Magnetic order and energy-scale hierarchy in artificial spin-ice structures},
  author = {Stopfel, H. and \"Ostman, E and Chioar, I-A and Greving, D and Arnalds, U. B. and Hase, T. P. A. and Stein, A. and Hj\"orvarsson, B. and Kapaklis, V.},
  journal = {Phys. Rev. B},
  volume = {98},
  pages = {014435},
  year = {2018},
  url = {https://doi.org/10.1103/PhysRevB.98.014435}
}

@article{Wang2006,
    author = {Wang, R. F. and Nisoli, C. and Freitas, R. S. and Li, J. and McConville, W. and Cooley, B. J. and  Lund, M. S. and Samarth, N. and  Leighton, C. and Crespi, V. H. and Schiffer, P.},
    title = {Artificial ‘spin ice’ in a geometrically frustrated lattice of nanoscale ferromagnetic islands},
    journal = {Nature},
    volume = {439},
    pages = {303-306},
    year = {2006},
    url = {https://doi.org/10.1038/nature04447},
}

@article{fiocco2014,
  title = {Encoding of {{Memory}} in {{Sheared Amorphous Solids}}},
  author = {Fiocco, Davide and Foffi, Giuseppe and Sastry, Srikanth},
  year = 2014,
  journal = {Phys. Rev. Lett.},
  volume = {112},
  pages = {025702},
  url = {https://doi.org/10.1103/PhysRevLett.112.025702}
}

@article{keim2020,
  title = {Global Memory from Local Hysteresis in an Amorphous Solid},
  author = {Keim, Nathan C. and Hass, Jacob and Kroger, Brian and Wieker, Devin},
  year = 2020,
  journal = {Phys. Rev. Research},
  volume = {2},
  pages = {012004},
  url = {https://doi.org/10.1103/PhysRevResearch.2.012004}
}

@article{mukherji2019,
  title = {Strength of {{Mechanical Memories}} Is {{Maximal}} at the {{Yield Point}} of a {{Soft Glass}}},
  author = {Mukherji, Srimayee and Kandula, Neelima and Sood, A. K. and Ganapathy, Rajesh},
  year = 2019,
  journal = {Phys. Rev. Lett.},
  volume = {122},
  pages = {158001},
  url = {https://doi.org/10.1103/PhysRevLett.122.158001}
}

@article{jules2022,
  title = {Delicate Memory Structure of Origami Switches},
  author = {Jules, Th{\'e}o and Reid, Austin and Daniels, Karen E. and Mungan, Muhittin and Lechenault, Fr{\'e}d{\'e}ric},
  year = 2022,
  journal = {Phys. Rev. Research},
  volume = {4},
  pages = {013128},
  url = {https://doi.org/10.1103/PhysRevResearch.4.013128}
}

@article{keim2019,
  title = {Memory Formation in Matter},
  author = {Keim, Nathan C. and Paulsen, Joseph D. and Zeravcic, Zorana and Sastry, Srikanth and Nagel, Sidney R.},
  year = 2019,
  journal = {Rev. Mod. Phys.},
  volume = {91},
  pages = {035002},
  url = {https://doi.org/10.1103/RevModPhys.91.035002}
}

@article{mungan2019b,
  title = {Networks and {{Hierarchies}}: {{How Amorphous Materials Learn}} to {{Remember}}},
  author = {Mungan, Muhittin and Sastry, Srikanth and Dahmen, Karin and Regev, Ido},
  year = 2019,
  journal = {Phys. Rev. Lett.},
  volume = {123},
  pages = {178002},
  url = {https://doi.org/10.1103/PhysRevLett.123.178002}
}

@article{mungan2025,
  title = {Self-{{Organization}} and {{Memory}} in a {{Disordered Solid Subject}} to {{Random Driving}}},
  author = {Mungan, Muhittin and Kumar, Dheeraj and Patinet, Sylvain and Vandembroucq, Damien},
  year = 2025,
  journal = {Phys. Rev. Lett.},
  volume = {134},
  pages = {178203},
  url = {https://doi.org/10.1103/PhysRevLett.134.178203}
}

@article{mungan2025a,
  title = {Self-Organization, {{Memory}} and {{Learning}}: {{From Driven Disordered Systems}} to {{Living Matter}}},
  author = {Mungan, Muhittin and Clement, Eric and Vandembroucq, Damien and Sastry, Srikanth},
  year = 2025,
  journal = {arXiv:2510.25367},
  url = {https://doi.org/10.48550/arXiv.2510.25367}
}

@article{oguz2020c,
  title = {Topology {{Restricts Quasidegeneracy}} in {{Sheared Square Colloidal Ice}}},
  author = {O{\u g}uz, Erdal C. and {Ortiz-Ambriz}, Antonio and {Shem-Tov}, Hadas and {Babi{\`a}-Soler}, Eric and Tierno, Pietro and Shokef, Yair},
  year = 2020,
  journal = {Phys. Rev. Lett.},
  volume = {124},
  pages = {238003},
  url = {https://doi.org/10.1103/PhysRevLett.124.238003}
}

@article{paulsen2025a,
  title = {Mechanical {{Memories}} in {{Solids}}, from {{Disorder}} to {{Design}}},
  author = {Paulsen, Joseph D. and Keim, Nathan C.},
  year = 2025,
  journal = {Annu. Rev. Condens. Matter Phys.},
  volume = {16},
  pages = {61--81},
  url = {https://doi.org/10.1146/annurev-conmatphys-032822-035544}
}

@article{sultana2025,
  title = {Ice Sculpting: {{An}} Artificial Spin Ice {{Tutorial}} on Controlling Microstate and Geometry for Magnonics and Neuromorphic Computing},
  author = {Sultana, Rawnak and Mondal, Amrit Kumar and Bhat, Vinayak Shantaram and Stenning, Kilian and Li, Yue and Arroo, Daan M. and Vasdev, Aastha and McCarter, Margaret R. and De Long, Lance E. and Hastings, J. Todd and Gartside, Jack C. and Jungfleisch, M. Benjamin},
  year = 2025,
  journal = {J. Appl. Phys.},
  volume = {138},
  pages = {061101},
  url = {https://doi.org/10.1063/5.0274799}
}

@article{meeussen_NJP_2020,
title = {Response evolution of mechanical metamaterials under architectural transformations},
volume = {22},
url = {https://doi.org/10.1088/1367-2630/ab69b5},
journal = {New J. Phys.},
author = {Meeussen, Anne S. and Oğuz, Erdal C. and van Hecke, Martin and Shokef, Yair},
year = {2020},
pages = {023030},
}

@article{coulais2016metacube,
title={Combinatorial design of textured mechanical metamaterials},
author={Coulais, Corentin and Teomy, Eial and de Reus, Koen and Shokef, Yair and van Hecke, Martin},
journal={Nature},
volume={535},
pages={529},
year={2016},
url = {https://doi.org/10.1038/nature18960}
}

@article{pisanty2021,
title={Putting a spin on metamaterials: {Mechanical} incompatibility as magnetic frustration},
author={Pisanty, Ben and O{\u{g}}uz, Erdal C. and Nisoli, Cristiano and Shokef, Yair},
journal={SciPost Phys.},
volume={10},
pages={136},
year={2021},
url = {https://doi.org/10.21468/SciPostPhys.10.6.136}
}
\end{document}